\documentclass[12pt]{article}
\usepackage{amsmath, amsthm}
\usepackage{multirow}
\usepackage{amsfonts}
\usepackage{amssymb,graphics,psfrag}
\usepackage{array,epsfig,multirow,stmaryrd,graphicx}
\usepackage{comment}
\numberwithin{equation}{section}
\numberwithin{table}{section}

\def\2{{1\over2}}
\let\<=\langle \let\>=\rangle
\def\new#1\endnew{{\bf #1}}
\def\ifundefined#1{\expandafter\ifx\csname#1\endcsname\relax}
\ifundefined{draftmode}\else    \input draftmode        \fi
\let\Msize=\footnotesize             
\def\BM{\Msize\begin{matrix}}           \def\EM{\end{matrix}}
\def\MN M:#1 #2 N:#3 #4 {{(#1_{#2},#3_{#4})}}
\def\MNH M:#1 #2 N:#3 #4 H:#5,#6 [#7]{{(#1_{#2},#3_{#4})^{#5,#6}_{#7}}}

\newcommand{\tr}{{\rm Tr}}

\newcommand{\sectiono}[1]{\section{#1}\setcounter{equation}{0}}

\def\CF{{\cal F}}

\def\CM{{\cal M}}
\def\CN{{\cal N}}
\def\CO{{\cal O}}
\def\CP{{\cal P}}
\def\CQ{{\cal Q}}

\newcommand{\be}{\begin{equation}}
\newcommand{\ee}{\end{equation}}
\newcommand{\bea}{\begin{eqnarray}}
\newcommand{\eea}{\end{eqnarray}}

\def\IZ{{\mathbb Z}}
\def\IR{{\mathbb R}}

\def\IP{{\mathbb P}}
\def\IT{{\mathbb T}}
\def\IS{{\mathbb S}}

\newcommand{\re}{{\rm e}}
\newcommand{\ri}{{\rm i}}
\newcommand{\rd}{{\rm d}}

\newcommand{\ba}{\begin{aligned}}
\newcommand{\ea}{\end{aligned}}
\newcommand{\ben}{\begin{eqnarray}\displaystyle}
\newcommand{\een}{\end{eqnarray}}

\def\tr{\hbox{tr}}

\newcommand{\figref}[1]{Fig.~\protect\ref{#1}}

\def\blfootnote{\xdef\@thefnmark{}\@footnotetext}
\long\def\symbolfootnote[#1]#2{\begingroup%
\def\thefootnote{\fnsymbol{footnote}}\footnote[#1]{#2}\endgroup}

\begin{document}

\begin{titlepage}

\hfill\vbox{
\hbox{MAD-TH-07-05}
\hbox{CERN-PH-TH/2007-070}
}

\vspace*{ 2cm}

\centerline{\Large \bf Black Holes and Large Order Quantum Geometry}

\medskip

\vspace*{4.0ex}

\centerline{\large \rm
Min-xin Huang$^{a}$, Albrecht Klemm$^{a,b}$, Marcos Mari\~no$^{c}$ and Alireza Tavanfar$^{c,d}$}

\vspace*{4.0ex}
\begin{center}
{\em $^a$Department of Physics and $^b$Department of Mathematics, \\[.1 cm]
University of Wisconsin, Madison, WI 53706, USA}

\vspace*{1.8ex}
{\em $^c$Department of Physics, CERN\\[.1cm]
Geneva 23, CH-1211 Switzerland}

\vspace*{1.8ex}

{\em $^d$Institute for Studies in Theoretical Physics and Mathematics (IPM) \\[.1cm]
P.O. Box 19395-5531, Tehran, Iran}

\symbolfootnote[0]{\tt minxin@physics.wisc.edu,\
aklemm@physics.wisc.edu, \, marcos@mail.cern.ch, \,
alireza.tavanfar@cern.ch}

\vskip 0.5cm
\end{center}

\centerline{\bf Abstract}
\medskip
We study five-dimensional black holes obtained by compactifying
M theory on Calabi--Yau threefolds. Recent progress in solving
topological string theory on compact, one-parameter models
allows us to test numerically various conjectures about these black holes. We
give convincing evidence that a microscopic description based on Gopakumar--Vafa invariants
accounts correctly for their macroscopic entropy, and we check that highly nontrivial cancellations --which seem
necessary to resolve the so-called entropy enigma in the OSV conjecture-- do in
fact occur. We also study analytically small 5d black holes obtained by wrapping M2 branes in the
fiber of K3 fibrations. By using heterotic/type II duality we obtain
exact formulae for the microscopic degeneracies in various geometries, and we compute their asymptotic
expansion for large charges.
\vskip 1cm

\noindent April 2007
\end{titlepage}

\tableofcontents

\section{Introduction}

String theory can provide in many situations a precise microscopic description of supersymmetric
black holes which reproduces for large charges the Bekenstein--Hawking entropy \cite{sv}. 
Degeneracies of microstates that are highly protected by supersymmetry are often
counted by mathematically well understood quasitopological quantities related to the compactification 
manifold. For example, the computation of microstates of the D1--D5 system is encoded in the elliptic 
genus of a symmetric product of a hyperK\"ahler manifold (see \cite{d,david} for a review of these computations).

A very challenging class of black holes in string theory is obtained by compactifying M theory on a 
Calabi--Yau manifold $X$ with generic $SU(3)$ holonomy. These are five dimensional black holes, which 
are characterized by a membrane charge $Q \in H_2(X,\IZ)$ and an angular momentum $m$. It 
was proposed in \cite{kkv} that the microscopic entropy of these black holes is accounted for 
by BPS states associated to M2 branes wrapping the cycle $Q$ and with left spin $m=2j_L$ in five dimensions. 
Accoording to the proposal of \cite{kkv} their degeneracies are encoded in the five dimensional supersymmetric index 
\begin{equation}
I(\alpha,\beta)={\rm Tr}(-1)^{2 j_L+2 j_R} \exp(-\alpha j_L- \beta H).
\label{5dindex} 
\end{equation} 
The information captured by this index can be extracted from the all-genus expansion of 
the holomorphic free energy of the topological string, computed at the large radius point of $X$ \cite{gv},
\begin{equation} 
\lim_{\bar t\rightarrow \infty} F(t,\bar t,g_s)=\sum_{g=0}^\infty g_s^{2g-2} F_g(t).
\label{freeenergy} 
\end{equation} 
In this identification 
the BPS degeneracies are mapped to the Gromov--Witten invariants of genus $g$ holomorphic 
maps. Since the topological string could not be solved on compact Calabi-Yau threefolds 
at higher genus, progress in understanding the microscopic degeneracies in 
the general case was very limited. On the boundary of the K\"ahler cone the 
problem might reduce effectively to a counting problem on a two complex dimensional 
surface, which is mathematically simpler, but the situation is also physically  
more degenerate. When the compactification manifold is $X=X_2 \times \IT^2$, where $X_2=\IT^4$ or K3, 
one obtains the five dimensional black hole solutions constructed in \cite{bmpv}, and 
the microscopic degeneracies are encoded in the elliptic genus of symmetric products of $X_2$.

In this paper we study the microscopic counting proposed in \cite{kkv} in two different situations, by using 
numerical and analytic methods. First of all, we consider 5d black holes obtained by compactifying M--theory on 
the one--parameter Calabi--Yau spaces studied in \cite{hkq}. This explores a generic direction in the 
K\"ahler cone and allows to describe generic 5d black holes, which have non-vanishing classical horizon area and can 
carry spin. In \cite{hkq} significant progress was made in solving the topological string on compact Calabi-Yau threefolds. 
By combining the holomorphic anomaly of \cite{bcov} with modularity properties of the topological string 
partition function $Z=\exp(F)$, effective action arguments, and Castelnouvo theory, it was possible 
to calculate the topological string free energy up to genus $53$.

In order to make contact with black hole physics on the (super)gravity side, one has
to obtain the asymptotic expansion of the microscopic degeneracies for large charge $Q$ 
and $Q\gg m$. For fixed $g$ the expansion of $F_g(t)$ around large radius is 
convergent and under analytic control by mirror symmetry. In contrast, the 
genus expansion in (\ref{freeenergy}) is expected to be asymptotic, as in noncritical 
string theories \cite{shenker} (see \cite{order} for a recent discussion of this issue in the 
context of topological strings). To obtain a large $Q$ expansion for the degeneracies of the 
$(Q,m)$ states one needs information at genus $g\sim Q^2$ and is hence 
facing the issues of the behavior of string theory at large genus. 
Although we don't have enough control of the degeneracies to obtain analytical
results on the large charge expansion, the situation is suited to a 
numerical study by using the method of Richardson transforms\footnote{For sub-sub leading 
terms we use the Pad\'e approximation.}. This method 
merely relies on the knowledge of the form of the series and makes it possible to 
extract its coefficients from the value of the degeneracies at finitely many points. 
The analysis is complicated by the fact that the 
large charge expansion of the degeneracies is an asymptotic expansion, but 
we find that the Richardson transforms converge rapidly to the expected macroscopic values for the asymptotic
coefficients. To estimate its accuracy
we sample over the thirteen Calabi-Yau, which have a
sufficently wide variety of topological data. Using this sample we
can conclude that, given our present data of the higher genus instanton expansion,
the leading coefficient of the asymptotic expansion is correct within
$2\%$ and the first subleading one within $12\%$. With this information at hand, we 
give convincing evidence that the topological string accounts correctly 
for the entropy of 5d spinning black holes, as conjectured in \cite{kkv}\footnote{For a 
recent study of this question by using an approach totally different from ours, see \cite{Guica:2007gm}.}.

Some aspects of the genus expansion (\ref{freeenergy}) are much 
better understood in terms of D-brane invariants like Gopakumar--Vafa or 
Donaldson--Thomas invariants, rather than Gromov--Witten invariants. 
In particular, for a given charge $Q\in H_2(X,\mathbb{Z})$ and $Q\neq 0$ 
one gets the complete genus information from a finite number 
of GV invariants. We use the results for $F_g$ in \cite{hkq} to obtain precise 
information on the Donaldson--Thomas invariants of the one-parameter models. This allows 
us to study numerically the scaling exponent $k$ considered in \cite{dm} (and defined below 
in (\ref{dtscaling}) and (\ref{limit})), 
which governs the growth 
of the Donaldson--Thomas invariants under rescalings of the charges. Our numerical 
study indicates that $k=2$. As argued in \cite{dm}, this value indicates that  highly nontrivial 
cancellations occur between the contributions to the Donaldson--Thomas invariants, which in turn 
seem necessary to resolve the so called entropy 
enigma \cite{dm} in the OSV conjecture \cite{osv}.

The second class of black holes we study has a different flavor. These are
5d black holes which are obtained when the Calabi--Yau is a K3 fibration and the charge $Q$ is restricted to the
K3 fiber. Their classical horizon area is zero (i.e. they are small black holes) and have no spin. By using
heterotic/type II duality one
can obtain analytic formulae for the $F_g$ amplitudes at all $g$ \cite{agnt,mm,kkrs,km,gkmw},
and from them one can extract the exact microscopic degeneracies for the
corresponding small 5d black holes. Of course,
as explained for example in \cite{dm}, the most delicate aspects of
5d spinning black holes, as well as of the OSV conjecture,
cannot be tested with small black holes. This reflects the fact that the Gromov--Witten theory of K3 fibers
(which is closely related to the theory of Hilbert schemes) is much simpler than the
Gromov--Witten theory of generic Calabi--Yau manifolds. However, having an exact
microscopic counting might be important in understanding some detailed aspects of the entropy.  As in the 4d case
considered in \cite{ddmp}, the 5d degeneracies are closely related to modular forms, but
one cannot use the Rademacher expansion featured in \cite{dmmv,ddmp}. We find however
an exact asymptotic expansion for the microscopic degeneracies in powers of the inverse charge (albeit
corrected by terms which are exponentially suppressed for large charges). The leading
term of the asymptotics is in agreement with a macroscopic computation using the 4d/5d
connection of \cite{gsy} and the 4d attractor equations for a D6/D2 system.

The organization of this paper is as follows. In section 2 we review the macroscopic and microscopic computation of the
entropy for 5d spinning black holes. In section 3 we analyze numerically the asymptotic properties of the
degeneracies for the Calabi--Yau manifolds studied in \cite{hkq}. 
In section 4 we study the asymptotic properties of Donaldson--Thomas invariants to address 
the entropy enigma of \cite{dm}. In section 5 we study small black holes in K3 fibrations and compute their degeneracies
as well as the asymptotic expansion. Finally, in section 6 we list some conclusions and open problems.

\section{Microscopic and macroscopic counting for 5d black holes}

\subsection{Macroscopic description}

Let us start with the macroscopic description of black hole
entropy. We will consider 5d black holes obtained by compactifying M theory on
a Calabi--Yau threefold $X$, and characterized by a charge $Q\in H_2(X,\IZ)$ and
$SU(2)_L\subset SO(4)$ angular momentum $m$. We will introduce a basis $\Sigma^A$ for $H_2(X,\IZ)$, where
$A=1, \cdots, b_2(X)$, as well
as a dual basis $\eta_A$ for $H^2(X)$. With respect to the $\Sigma^A$ basis, the charge $Q$
will be given by a set of integers $Q_A$.
The classical entropy of these black holes, denoted as $S_0$, is one
quarter of the horizon area
\begin{eqnarray} \label{entropy1}
S_0= 2\pi\sqrt{\CQ^3-m^2},
\end{eqnarray}
where $\CQ$ is the graviphoton charge of the black hole.
This charge is related to the membrane charge $Q$ by the
attractor mechanism in five dimensions \cite{Ferrara:1996dd}, which states that
\be
\CQ^{3/2} =\frac{1}{6}C_{ABC}y^A y^B y^C,
\ee
where
\be
{1\over 2} C_{ABC} y^B y^C =Q_A,
\ee
and
\be
C_{ABC}=\int_X \eta_A \wedge \eta_B \wedge \eta_C
\ee
 are the triple intersection numbers of $X$. For one-parameter models, the
membrane charge will be identified with the degree $d$ of
the holomorphic map in topological string computations, and we will denote the single
intersection number by $C_{ABC}=\kappa$. From the above equations it follows that
\be \label{graviphoton-03-17}
\CQ=\biggl(\frac{2}{9\kappa}\biggr)^{\frac{1}{3}}d.
\ee

There is a correction to the black hole entropy from
the $R^2$ term of the supergravity effective action, which we
denote as $S_1$ for convenience. The $R^2$ term correction to the
black hole entropy scales like $\CQ^{\frac{1}{2}}$ in the large charge
limit, and was computed in \cite{Guica:2005ig} by using Wald's
formula \cite{wald} for the $R^2$ in 5d. The result reads,
\begin{eqnarray} \label{entropy-03-17-01}
S_1=\frac{\pi}{24}\sqrt{Q^3-m^2}~c_2\cdot
Y\biggl(\frac{3}{Q}+\frac{m^2}{Q^4}\biggr)
\end{eqnarray}
where
\be
Y^A ={1\over \CQ^{1/2}} y^A,
\ee
and
\be
\label{ctwo}
c_{2A} =\int_{X} c_2(X) \wedge \eta_A.
\ee
For $m=0$ this formula has been rederived in \cite{Castro:2007hc,Alishahiha:2007nn} by using
the full 5d SUGRA action. In the one-parameter case, this correction reads
\begin{eqnarray} \label{entropy-03-17-02}
S_1=\frac{\pi
c_2}{8}\biggl(\frac{6}{\kappa}\biggr)^{\frac{1}{3}}~\sqrt{Q^3-m^2}
\biggl(\frac{1}{Q}+\frac{m^2}{3Q^4}\biggr)
\end{eqnarray}

Besides the corrections that we have considered, there are well known correction terms in the low
energy effective action of the form $F^{2g-2}R^2$, $g\geq 2$,
where $F$ is the graviphoton field strength. The leading
contribution comes from a classical term, which is the
contribution of the constant map from a genus $g$ world-sheet to
the Calabi-Yau manifold. It is of the form, 
\be
\label{constantmap-03-17}
d_g \chi
\ee
where $\chi$ is the Euler number of the Calabi-Yau 3-fold and 
\begin{equation} 
\label{dg}
d_g=\frac{(-1)^g |B_{2g}B_{2g-2}|}{4g(2g-2)(2g-2)!}.
\end{equation}
We denote the correction to black hole entropy due to the
$F^{2g-2}R^2$ term as $S_g$. We can roughly estimate the
correction for non-spinning black holes $m=0$ as follows.

The graviphoton charge is the integral of its field strength over
the horizon of black hole,
\begin{eqnarray}
\CQ \sim \int_{\textrm{Horizon}}F
\end{eqnarray}
Since the area of the horizon scales like $A\sim \CQ^{\frac{3}{2}}$, the
graviphoton field strength goes like
\begin{eqnarray}
\label{FQ}
F\sim \CQ^{-\frac{1}{2}}
\end{eqnarray}
The $R^2$ term contributes a factor $\CQ^{-1}$ in Wald's formula
for the black hole entropy, and taking into account also the factor of
horizon area $A\sim \CQ^{\frac{3}{2}}$, we find the scaling behavior
of the $F^{2g-2}R^2$ term correction to black hole entropy to be
\begin{eqnarray} \label{highergenus-03-17}
S_g\sim \chi \CQ^{\frac{3}{2}-g}
\end{eqnarray}
where we have included the Euler number from
(\ref{constantmap-03-17}). The constant of proportionality in
(\ref{highergenus-03-17}) is now universal, and independent of
specific Calabi-Yau geometries and the black hole charge. We will
be able to make a rough test of (\ref{highergenus-03-17}) for the
genus $2$ case, which is the sub-sub-leading correction in the large
$\CQ$ limit.

There are other worldsheet instanton corrections to the low
energy effective action that can be computed also by topological
strings. However, these terms are exponentially small in large
charge $\CQ$ limit where the supergravity description is valid, and
are much suppressed compared to the $\CQ^{-1}$ power corrections
 in (\ref{highergenus-03-17}). In this paper we will not need
to consider these world-sheet instanton corrections in the
macroscopic description of the black hole entropy. Interestingly
these world-sheet instanton corrections are closely related to the BPS
states that we will count in the microscopic description of the
black hole entropy.

\subsection{Microscopic description}

Microscopically, a 5d
black hole with
membrane charge $Q \in H_2(X,\IZ)$ is engineered by wrapping M2 branes around the two--cycle $Q$.
This leads to a supersymmetric spectrum of BPS states in 5d
which are labeled by $Q$ and by their spin content $(j_L, j_R)$. As argued in
\cite{gv}, in order to obtain an index one has to
trace over $j_R$ (with an insertion of $(-1)^{2 j_R}$). The resulting spectrum for a membrane charge $Q$ can be
represented as
\be
\label{repQ}
R_Q=\sum_{r=0}^g n_Q^{r}I_{r+1}
\ee
where
\be
I_{\ell}=\biggl[ 2({\bf 0}) + \biggl( {\bf 1\over 2}\biggr)\biggr]^{\ell}
\ee
encodes the spin content $j_L$, and $n_Q^r$ are the Gopakumar--Vafa invariants \cite{gv}.
Notice that in (\ref{repQ}), the sum over $r$ is finite and the highest spin $g$ appearing
in the sum depends on the membrane charge $Q$. In other words, for any given
$Q$ there are only finitely many $g$ for which the $n_Q^r$ are nonzero.

We can now write down a generating
function for the supersymmetric degeneracies of BPS states with membrane charge $Q$, keeping
track of their left spin $j_L$, by computing
\be
\sum_m \Omega(Q,m)= \sum_Q \tr_{R_Q} (-1)^{2j_L}y^{j_L}.
\ee
Using the decomposition (\ref{repQ}) one finds
\be
\label{degs}
\Omega(Q,m) = \sum_{r}  {2r+2 \choose m+r+1} n_Q^r.
\ee
where $m=2j_L$. In \cite{kkv} it was proposed that this quantity
gives the microscopic degeneracies for a spinning 5d black hole of charge $Q$ and spin $J=m$. The computation
of these degeneracies reduces then to the computation of the Gopakumar--Vafa invariants $n_Q^r$. The most effective way to determine these is
by computing the genus $r$ amplitudes $F_r$ of topological string theory on $X$. As shown in \cite{gv},
the total free energy
\be
\label{totalf}
F(t,g_s)=\sum_{r=0}^{\infty} F_r (t) g_s^{2r-2}
\ee
can be written in terms of the Gopakumar--Vafa invariants as
\be
F(t,g_s)=\sum_{r=0}^\infty  \sum_{Q \in H_2(X)} \sum_{k=1}^{\infty}
n^r_Q {1\over k}
\left(2 \sin {k g_s \over 2}\right)^{2 r-2}  {\rm e}^{-k Q\cdot t}.
\label{gova}
\ee
This means, in particular, that one can obtain the $n^r_Q$ for all $Q$ by knowing $F_0, \cdots, F_r$.
The black hole entropy is given by  logarithm of the number of microstates
\begin{eqnarray}
S(Q,m)=\log(\Omega(Q,m)).
\end{eqnarray}
This should agree with the macroscopic result in the large charge limit $Q\gg 1$ and $Q\gg m$.

As explained in \cite{kkv}, this proposal for the microscopic counting of states of 5d black holes can be regarded as a generalization of the
elliptic genus, which computes BPS degeneracies of the D1--D5 system. Indeed, if one considers M theory compactified on $X={\rm K3} \times \IT^2$,
the generic M2 brane charge in this compactification is
\be
\label{prodcharge}
Q=[C] + n [\IT^2], \quad n \in \IZ,
\ee
and $C$ is a curve in K3. By standard dualities this system can be related to type IIB on K3$\times \IS^1$ with D--brane charge $[C]$ and $M$ units
of momentum around $\IS^1$, which is a close cousin of the D1--D5 system.
As it is well known (see for example \cite{d}), the BPS degeneracies of this system can be computed from the
elliptic genus of the symmetric product of K3. Let
\be
\label{egenera}
\chi({\rm K3};q,y)=\sum_{m,\ell}c(m,\ell) q^m y^{\ell}
\ee
be the elliptic genus of K3. The generating function of elliptic genera of the symmetric product $S^k {\rm K3}$
\be
\label{egsym}
\chi(S_p {\rm K3};q,y) =\sum_{k=0}^p \chi (S^k {\rm K3};q,y)p^k =\sum_{k,n,m} c(k,n,m) p^k q^n y^m
\ee
can be computed from the coefficients in (\ref{egenera}) in terms of an infinite product
\cite{dmvv}
\be
\label{egip}
\chi(S_p {\rm K3};q,y)= \prod_{N,M\ge 0, \ell} (1-p^N q^M y^{\ell})^{-c(N M, \ell)}.
\ee
 The supersymmetric degeneracies of  BPS states for this system are then given by the coefficients of the
 expansion in (\ref{egsym}),
 \be
 \label{egcounting}
 \Omega(Q,m)=c\Bigl({1\over 2} C^2 +1, n, m\Bigr),
 \ee
where $Q$ is of the form (\ref{prodcharge}). One can show that, for large charges \cite{dvv,d},
\be
\log \, \Omega(Q,m) \sim 2\pi {\sqrt { {n\over 2} C^2  -m^2}}.
\ee
It is easy to check that this is precisely the macroscopic entropy (\ref{entropy1}) computed for K3$\times \IT^2$.
Of course, the degeneracies (\ref{degs}) are in general much more difficult to compute, since they correspond to
black holes with only $\CN=1$ supersymmetry in 5d.

\sectiono{One--parameter models}

\subsection{Topological strings on one--parameter models}

In \cite{hkq} the topological B model was integrated on thirteen one-parameter
Calabi-Yau spaces which can be realized as
hypersurfaces or complete intersections in (weighted) projective
spaces. We have listed these spaces and some of their topological data in table \ref{onep}.
These data are the intersection numbers $C_{ABC}=\kappa$, the second
Chern classes $c_2$, and the Euler numbers $\chi$. They are needed for
computations of the macroscopic entropy.

\begin{table} [t] \label{table1}
\begin{centering}
\begin{tabular}{|r|rrr||r|rrr|}
\hline CY & $\chi$& $c_2\cdot \eta$ & $\kappa$& CY          & $\chi$& $c_2\cdot \omega$ & $\kappa$\\   \hline
$X_5(1^5)$   &-200   &50                 &  5      & $X_6(1^4,2)$& -204&                 42 & 3 \\
$X_8(1^4,4)$  &-296   &44                 &  2      & $X_{10}(1^3,2,5)$& -288&            34 & 1 \\
$X_{3,3}(1^6)$&-144   &54                &  9      & $X_{4,2}(1^6)$& -176&               56 & 8 \\
$X_{3,2,2}(1^7)$&-144   &60              &  12     & $X_{2,2,2,2}(1^8)$    & -128&       64 & 16 \\
$X_{4,3}(1^5,2)$&-156   &48              &  6      & $X_{4,4}(1^4,2^2)$ & -144&          40 & 4 \\
$X_{6,2}(1^5,3)$&-256   &52              &  4      & $X_{6,4}(1^3,2^2,3)$& -156&         32 & 2 \\
$X_{6,6}(1^2,2^2,3^2)$&-120   &32        &  1      & & &                  &  \\
\hline
\end{tabular}
\caption{The sample of 13 one-parameter complete intersection CYs in weighted projective spaces.
A CICY of degree $d_1,\ldots, d_k$ in weighted projective space $\mathbb{P}^{l-1}(w_1,\ldots,w_l)$
is denoted $X_{d_1,\ldots, d_k}( w_1,\ldots,w_l)$, i.e. weights
$w$ with repetition $m$ are abbreviated by $w^m$. $\chi = \int_X c_3$ is the Euler number, $\kappa$ 
is the triple intersection number, and $c_2\cdot \eta $ is defined in (\ref{ctwo}).}
\end{centering}
\label{onep}
\end{table}

The complex moduli space of these threefolds is ${\cal M}=\mathbb{P}^1\setminus \{\infty,1,0\}$, and
the three special points are the large radius degeneration point, a conifold point and
a further point either of finite (Gepner point) or infinite branching order. The modular group
$\Gamma_X \in {\rm SP}(4,\mathbb{Z})$ can hence be generated e.g. by the large
radius and the conifold monodromies.

The  conceptual obstacle in integrating the B-model holomorphic anomaly~\cite{bcov}
is the holomorphic ambiguity which arises in each integration step.
Invariance of  the topological string amplitudes under $\Gamma_X$
and effective action arguments, which govern the behaviour of the genus
$g$ amplitudes at special points, restrict the ambiguity to
$3g-3$ unknowns~\cite{kkv}. By using a refined effective action analysis, which gives
rise to the ``gap condition" at the conifold, regularity at the orbifold, and Castelnouvo's bound for
the Gromow--Witten invariants at large radius, it is possible to fix the unknows,
and one can calculate the free energy of the topological
string to arbitrary degree and up to genus $12-53$.

Instead of using the generic solution of holomorphic anomaly equation
suggested by the world-sheet degenerations~\cite{bcov} we use the
constraints of $\Gamma_X$ on the topological string amplitudes directly 
when integrating the holomorphic anomaly equations genus by genus~\cite{Yamaguchi:2004bt}\cite{gkmw}. 
This results in an alogarithm, which constructs the genus $g$ amplitudes  
as weight $3g-3$ polynomials over a ring of three an-holomorphic- and one holomorphic 
modular objects of weight $(1,2,3,1)$. As a consequence the number of terms in the $F_g$
grows polynomial with $g$ and not exponentially as in the approach of~\cite{bcov}\footnote{Nevertheless 
the limiting factor in advancing to higher $g$ is presently not the ambiguity but the 
computing time. The reason is that the numerators in the coefficents of the 
polynomials grow exponentially.}.

The approach of \cite{hkq} views the topological string partition function as a wave function over
$H^3(X,\IR)$. Choices of polarization are necessary in order to expand the effective action
at different points in the moduli space $\CM$, in appropriate local holomorphic coordinates.
Most of the black hole issues that we will discuss involve the degeneracies
extracted from the topological string at the large radius limit.
Therefore we will discard for now most of the global
information and focus only the holomorphic limit of the topological string
partition function at this limit, where it encodes the degeneracy of
bound states of a single D6 brane and arbitrary D2-D0 branes.

\subsection{Static black holes}

We first consider the case of non-spinning black hole $J\equiv
m=0$ and denote $N_d=\Omega(d,0)$. The entropy formula including the
first few orders (\ref{entropy1}), (\ref{entropy-03-17-02}),
(\ref{highergenus-03-17}) is in this case
\begin{eqnarray} \label{entropy2}
S=b_0d^{\frac{3}{2}}+b_1d^{\frac{1}{2}}+\frac{b_2}{d^{\frac{1}{2}}}+\mathcal{O}\biggl(\frac{1}{d^{\frac{3}{2}}}\biggr),
\end{eqnarray}
where the first two coefficients are
\begin{eqnarray} \label{coefficients-03-17}
b_0=\frac{4\pi}{3\sqrt{2\kappa}},~~~~ b_1=\frac{\pi
c_2}{4\sqrt{2\kappa}},
\end{eqnarray}
and we have used the graviphoton charge relation
(\ref{graviphoton-03-17}). The coefficient $b_1$ of the
sub-leading term is consistent with the results in \cite{vafabh, Castro:2007hc, Alishahiha:2007nn},
where it was
observed that the $b_1$ can be produced by a shift of the 2-brane
charge
\be
d\rightarrow d+\frac{c_2}{8}
\ee
in the leading term.

To compare with the microscopic counting we define the following
function
\begin{eqnarray} \label{fd}
f(d)= \frac{\log(N_d)}{d^\frac{3}{2}}.
\end{eqnarray}
The macroscopic black hole entropy predicts that the large order
behavior of $f(d)$ is
\begin{eqnarray} \label{functionexpand-03-17}
f(d)=b_0+\frac{b_1}{d}+\frac{b_2}{d^2}+\cdots
\end{eqnarray}
Since we have available only the values of $f(d)$ for positive
integer $d$ up to a finite degree, it is appropriate to use
well-known numerical methods to extrapolate the asymptotic value
$b_0$. From the form of the sub-leading corrections in
(\ref{functionexpand-03-17}), it is appropriate to use the
Richardson extrapolation method (see for example \cite{bo}).

The basic idea of this numerical method is simple. To
cancel the sub-leading corrections in (\ref{functionexpand-03-17})
		   up to order $1/d^N$, one defines
\begin{eqnarray} \label{extra-03-17}
A(d,N)=\sum_{k=0}^{N}\frac{f(d+k)(d+k)^N(-1)^{k+N}}{k!(N-k)!},
\end{eqnarray}
One can show that if the perturbation series
(\ref{functionexpand-03-17}) truncates at order $1/d^N$,
the expression (\ref{extra-03-17}) will give exactly the
asymptotic value $b_0$. Ideally ,the larger $N$ and $d$ are, the
closer $A(d,N)$ is to the asymptotic value. But due to the
limitation of our data, the sum $d+N$ must not exceed the maximal
degree $d_{max}$ of the topological string computations.

\begin{figure}[hbtp]
\begin{center}
\epsfig{file=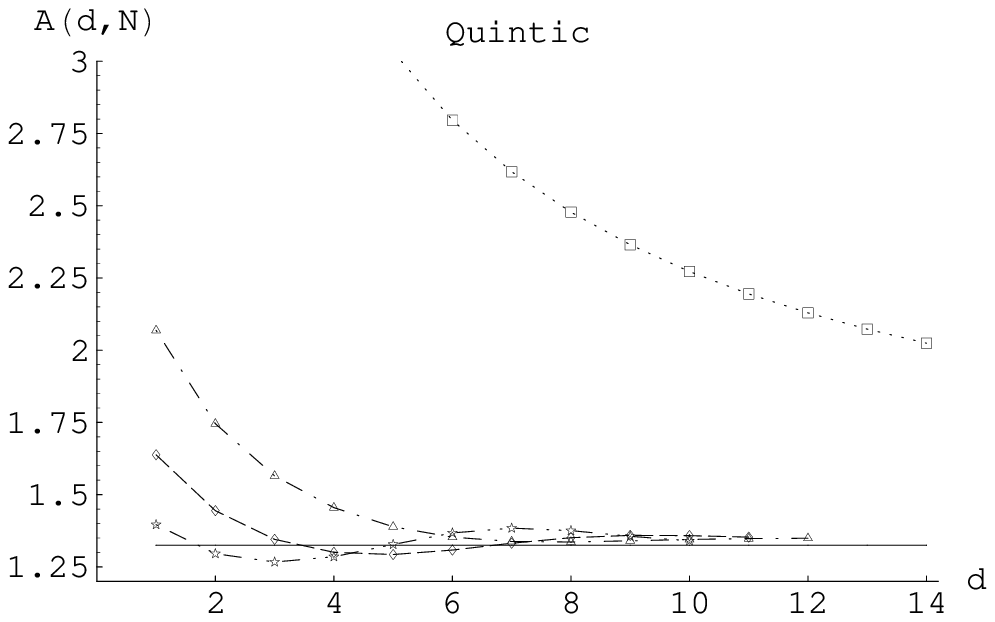, width=8cm}\epsfig{file=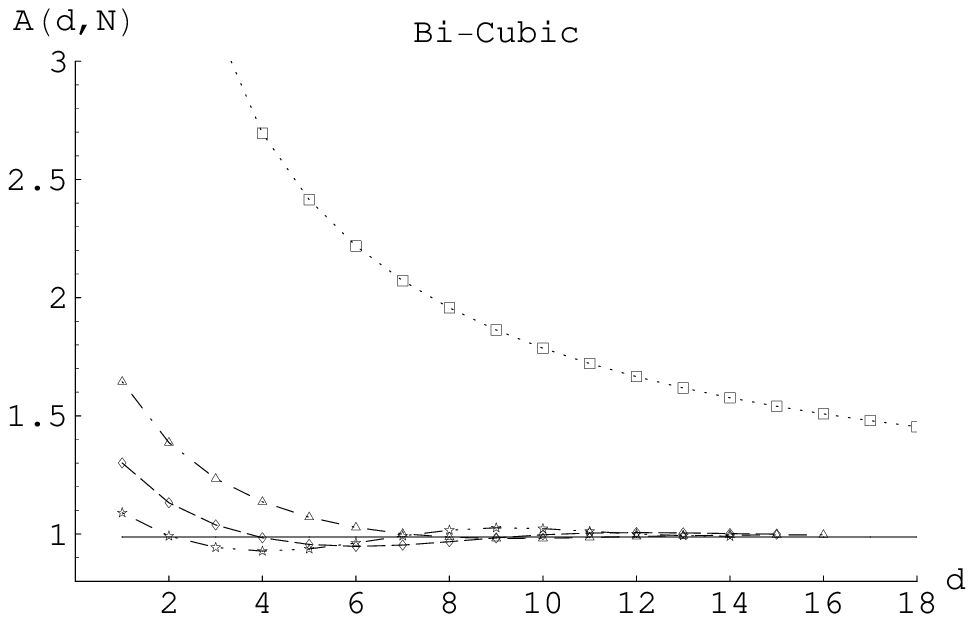, width=8cm}
\end{center}
\caption{Microscopic data for $f(d)$ ({\tiny{$\Box$}}), and the Richardson transforms  
$A(d,2)$ ({\tiny{$\triangle$}}), $A(d,3)$ ($\diamond$), and $A(d,4)$ ($\star$). 
The  straight line corresponds to the macroscopic prediction 
$b_0=\frac{4\pi}{3\sqrt{2\kappa}}$. For the quintic this value 
is $b_0\approx 1.359$ and for the available degree $14$ 
the Richardson transforms lie $1.8$, $2.1$, $1.2$ \% from the macroscopic  
prediction. For the bi-cubic $b_0\approx 0.967$, the available degree is higher, $18$, and the
microscopic counting is within $.9$, $1.2$, $.3$ \% from the macroscopic  
prediction.  As an example we give BPS numbers used for the analysis 
at degree $18$ of the bi-cubic in table (A.1).} 
\label{leading}
\end{figure}

Fig. \ref{leading} shows the convergence of the leading terms in  
$f(d)$ and of the Richardson transforms $A(d,N)$, $N=2,3,4$  
for the quintic  and the bi-cubic. It is obvious from the two examples 
in  Fig. \ref{leading} that the Richardson method 
improves impressively  the convergence of the series, i.e. it provides 
a model independent and consistent scheme to supress the subleading corrections.  
Using $N=2-4$ is good enough for our purpose of estimating the asymptotic 
value. We conduct the analysis for all 13 models using $N=3$ and 
the maximal degree available. The results are summarized in 
Table \ref{table3-03-17}, and are in very good agreements 
with the expected asymptotic values $b_0$ in
(\ref{coefficients-03-17}). More detailed results 
on all the analysis carried out in this paper can be found 
in a script and in a data base at \cite{webpage}.

\begin{table}
\begin{centering}
\begin{tabular}{|r|r|r|r|r|}
\hline Calabi-Yau &  $d_{max}$ & $A(d_{max}-3,3)$ & $b_0=\frac{4\pi}{3\sqrt{2\kappa}}$ & error \\
\hline  $X_{5}(1^5)$  & 14 & 1.35306 & 1.32461 &
2.15 \%  \\
 \hline   $X_{6}(1^4,2)$  & 10 & 1.75559 & 1.71007 & 2.66 \%
\\
\hline   $X_{8}(1^4,4)$  & 7 & 2.11454 & 2.0944 & 0.96 \%
\\
\hline  $X_{10}(1^3,2,5)$  & 5 & 2.99211 & 2.96192 & 1.02 \%
\\
\hline  $X_{3, 3}(1^6)$  & 17 & 1.00204 & 0.987307 & 1.49 \%   \\
\hline  $X_{4, 2}(1^6)$   & 15 & 1.07031 & 1.0472 & 2.21 \%  \\
\hline  $X_{3, 2, 2}(1^7)$  &  10 & 0.821169 & 0.855033 & -3.96 \% \\
\hline  $X_{2, 2, 2, 2}(1^8)$  & 13 & 0.722466 & 0.74048  &
-2.43 \%  \\
\hline  $X_{4, 3}(1^5,2)$  & 11 & 1.21626 & 1.2092 & 0.58  \%  \\
\hline  $X_{6, 2}(1^5, 3)$  & 11 & 1.52785 & 1.48096 & 3.17  \% \\
\hline  $X_{4, 4}(1^4,2^2)$  & 7 & 1.42401 & 1.48096 &
-3.85 \% \\
\hline  $X_{6, 4}(1^3,2^2, 3)$  & 5 & 2.06899 & 2.0944 & -1.21 \%
\\
\hline  $X_{6, 6}({1^2, 2^2, 3^2})$  & 4 & 2.95082 & 2.96192 &
-0.37 \% \\
 \hline
\end{tabular}  \caption{\label{table3-03-17} Comparing the extrapolated value of $b_0$ with the macroscopic prediction.}
\end{centering}
\end{table}

\begin{figure}[hbtp]
\begin{center}
\epsfig{file=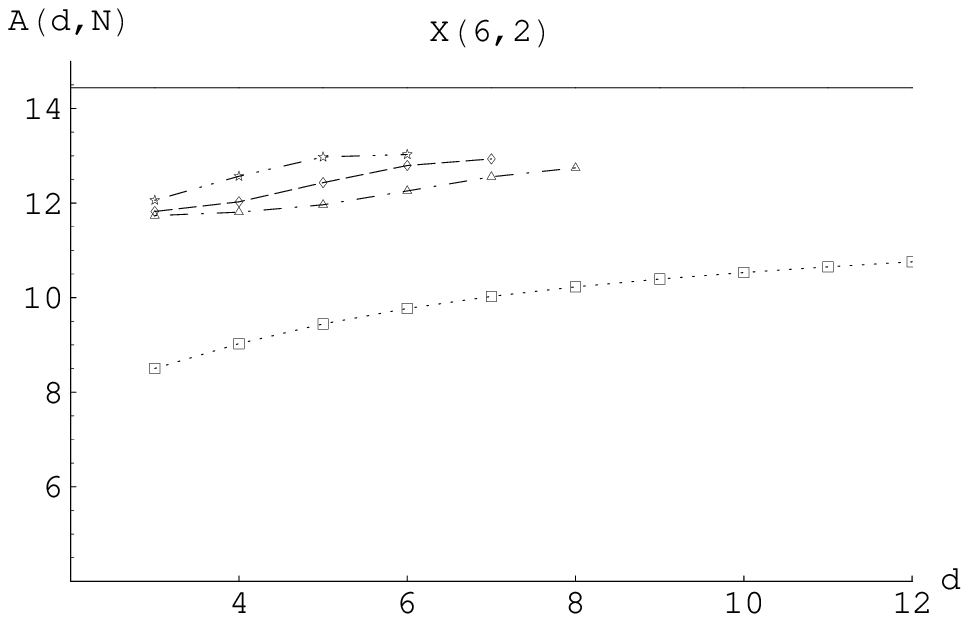, width=8cm}\epsfig{file=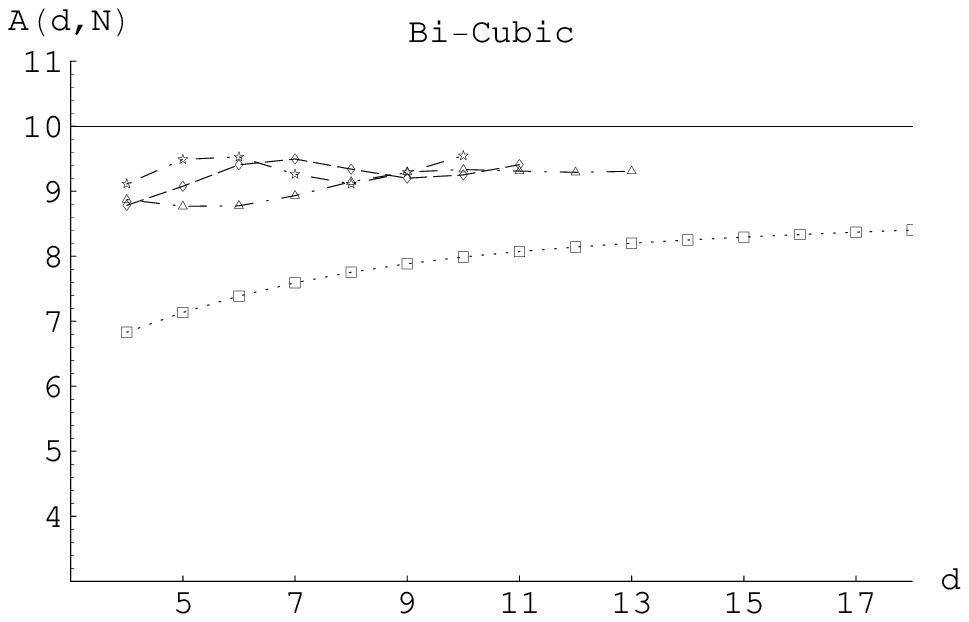, width=8cm}
\end{center}
\caption{
Microscopic data for $f(d)$ ({\tiny{$\Box$}}), and the Richardson transforms  
$A(d,4)$ ({\tiny{$\triangle$}}), $A(d,5)$ ($\diamond$), and $A(d,6)$ ($\star$).
The  straight line corresponds to the macroscopic prediction 
$b_1=\frac{\pi c_2}{4\sqrt{2\kappa}}$. 
For the degree $X_{6,2}$ complete intesection 
this value is $b_1\approx 14.44$ and for the available degree $12$ 
the Richardson transforms lie $-11.7$, $-10.4$, $-9.77$ \% below the macroscopic  
prediction. For the bi-cubic $b_1\approx 9.994$, the available degree is  $18$ and the
microscopic counting is  $-7.15$, $-6.88$, $-6.63$ \%  below the macroscopic  
prediction.}
\label{subleading}
\end{figure}

We can further extract the sub-leading coefficient $b_1$ from the
data. Define
\begin{eqnarray} \label{fd1}
f_1(d)&=& (f(d)-b_0)d ,\nonumber \\
A_1(d,N)&=&
\sum_{k=0}^{N}\frac{f_1(d+k)(d+k)^N(-1)^{k+N}}{k!(N-k)!},
\end{eqnarray}
and the asymptotic value of $f_1(d)$ should be $b_1$. We apply the
same Richardson extrapolation method to $f_1(d)$ and we compare it with
the macroscopic black hole predictions. Two typical examples for the 
behaviour of the Richardson transforms are plotted in Fig. \ref{subleading}.  
The results for all models are summarized in Table \ref{table4-03-17}.

\begin{table}
\begin{centering}
\begin{tabular}{|r|r|r|r|r|r|}
\hline Calabi-Yau & $d_{max}$  & $A_1(d_{max}-3,3)$ &
$b_1=\frac{\pi
c_2}{4\sqrt{2\kappa}}$ & error & estimated $b_2$ \\
\hline  $X_{5}(1^5)$  & 14 & 11.2668 & 12.4182 & -9.27 \%  & -11.9503  \\
\hline  $X_{6}(1^4,2)$  & 10 & 11.9237 & 13.4668 & -11.5 \%  & -12.1848  \\
\hline  $X_{8}(1^4,4)$  & 7 & 14.0537 & 17.2788 & -18.7 \%  & -14.9973  \\
\hline  $X_{10}(1^3,2, 5)$  &5 & 15.2509 & 18.8823 & -19.2 \%  & -14.9817  \\
\hline  $X_{3, 3}(1^6)$  & 17 & 9.29062 & 9.99649 & -7.06 \%  & -9.63958  \\
\hline  $X_{4, 2}(1^6)$  & 15 & 10.0226 & 10.9956 & -8.85 \%  & -10.7834  \\
\hline  $X_{3, 2, 2}(1^7)$  & 10 & 8.45163 & 9.61912 & -12.1 \%  & -9.3828   \\
\hline  $X_{2, 2, 2, 2}(1^8)$  & 13 & 7.84595 & 8.88577 & -11.7 \%  & -8.88773  \\
\hline  $X_{4, 3}(1^5, 2)$  & 11 & 9.5981 & 10.8828 & -11.8 \%  & -9.96404   \\
\hline  $X_{6, 2}(1^5, 3)$  & 11 & 12.5614 & 14.4394 & -13.0 \%  & -14.2582   \\
\hline  $X_{4, 4}({1^4, 2^2})$  & 7 & 9.70091 & 11.1072 & -12.7 \%  & -9.41295  \\
\hline  $X_{6, 4}(1^3, 2^2, 3)$  & 5 & 11.1008 & 12.5664 & -11.7 \%  & -10.0821   \\
\hline  $X_{6, 6}(1^1, 2^2, 3^3)$  & 4 & 11.1378 & 12.2179 & -8.84 \%  & -8.15739  \\
\hline
\end{tabular}  \caption{\label{table4-03-17} Comparing the extrapolated value with the macroscopic prediction of $b_1$.}
\end{centering}
\end{table}

Despite our rather successful verifications of the numerical
coefficients $b_0$ and $b_1$, we should note that the expansion in inverse powers of the charge
(\ref{functionexpand-03-17}) is actually an asymptotic series. The asymptotic character 
of the large charge expansions of microscopic degeneracies is manifest in the explicit computations for 
small black holes in \cite{ddmp} and also in the examples we will discuss in section 5. In our case, we can relate 
the asymptotic expansion of (\ref{functionexpand-03-17}) to a large genus behavior in a string series, since 
the coefficients in (\ref{functionexpand-03-17}) are 
proportional to the constant map contribution
\begin{eqnarray} \label{constant-03-29}
b_g\sim d_g,
\end{eqnarray}
where $d_g$ is given in (\ref{dg}). This coefficient grows at large $g$ as 
\be
d_g \sim (2\pi)^{-4g} (-1)^g (2g)!,
\ee
which is the typical behavior found in string perturbation theory \cite{shenker}.
It then follows that the series expansion (\ref{functionexpand-03-17}) for $f(d)$ 
has zero radius of convergence for any value of $d$ and it is 
rather an asymptotic expansion. Indeed, the $d_g$ are the coefficients of the asymptotic 
expansion of the MacMahon function (see \cite{ddmp}, Appendix E, for a detailed 
derivation). For these kinds of expansions, the best approximation to their true value (which in 
this case is the function $f(d)$ computed from topological strings) is obtained by 
truncating the sum at the order $\CN$ which minimizes the error. For an asymptotic series of the form
\be
\label{asympa}
f(w) =\sum_{k=1}^{\infty} b_k w^k, \qquad b_k \sim A^{-k} (\beta k)! 
\ee
the optimal truncation occurs generically at 
\be
\CN \sim \frac{1}{\beta} \biggl( \frac{A}{|w|}\biggr)^{\frac{1}{\beta}} .
\ee
In our case $\beta=2$ and we can estimate $\CN$ as follows. According to the connection between 4D/5D black holes \cite{gsy}, the attractor 
value for the topological string coupling constant is $g_s =4\pi$ \cite{Guica:2005ig}. This should 
be roughly the numerical constant that relates the graviphoton 
field strength to the charge $\CQ$ in (\ref{FQ}), and it contributes to the
coefficients $b_g$ an extra factor $g_s^{2g-2}$, so that we can refine (\ref{constant-03-29}) to
\be
b_g \sim (4 \pi)^{2g} d_g, 
\ee
and the constant in (\ref{asympa}) is $A=\pi^2$. Therefore, the optimal truncation is at 
\be
\CN \sim \frac{\pi}{2} d^{\frac{1}{2}} .
\ee
For the small values of $d$ that we are considering we should therefore expect 
an optimal truncation around $\CN \sim 5-10$. 

These considerations have implications for our numerical analysis. The Richardson method (\ref{extra-03-17}) is 
designed in principle for convergent expansions. For asymptotic expansions, we should expect it 
to give increasing precision and convergence to the true coefficients as long as the order of the 
transformation $N$ in (\ref{extra-03-17}) is lower than the truncation order $\CN$. This is the underlying reason that
prevents us from improving the precision of the leading
coefficients by simply increasing the truncation order $N$ in the
Richardson method, and we indeed find an oscillating behavior around the expected true 
value for the Richardson transforms with $N >5$.

We try to go one step further and give a rough estimation of the
coefficient $b_2$ in (\ref{entropy2}), which has not been studied
in the literature from the supergravity point of view. It turns
out that the naive method we use for computing the sub-leading
coefficient $b_1$ gives too big an estimate, which might be a
result that the optimal truncation scheme is no longer a good
approximation at this order. In order to improve this, we use the
Pad\'{e} approximation which is well-known for summing divergent
series. Given an asymptotic series
\begin{eqnarray} \label{originalseries-03-29}
f(z)=\sum_{i=0}^{\infty}b_iz^i, \end{eqnarray} one can evaluate
the asymptotic value by  defining the following Pad\'{e}
approximation
\begin{eqnarray} \label{pade-03-29}
P^{N}_M(z)=\frac{\sum_{i=0}^NA_iz^i}{1+\sum_{i=1}^MB_iz^i}
\end{eqnarray}
where the coefficients $A_i$ and $B_i$ are fixed by Taylor
expanding the above equation (\ref{pade-03-29}) around $z=0$ and
match to the first $M+N+1$ terms of the original series
(\ref{originalseries-03-29}).

We take the theoretical values of $b_0$ and $b_1$ from
(\ref{coefficients-03-17}), and use the Monte Carlo method to
randomly generate the sub-leading coefficients $b_2$, $b_3$ etc,
then use the Pad\'{e} approximation to evaluate the asymptotic
series (\ref{originalseries-03-29}) for $z=\frac{1}{d}$, where
$d=1,2,\cdots,d_{max}$. We pick the sub-leading coefficients $b_i$
($i\geq 2$) that minimize the difference of the Pad\'{e}
evaluation with the expected value $f(d)$  from topological
strings, i.e. we minimize
\begin{eqnarray}
\sum_{d=1}^{d_{max}}(\frac{P^N_M(\frac{1}{d})}{f(d)}-1)^2
\end{eqnarray}
We find different values of $N$, $M$ in the Pad\'{e} approximation
give qualitatively similar results. In the last column in Table
\ref{table4-03-17}, we give the estimated values of
sub-sub-leading coefficient $b_2$ using the scheme $N=2,M=1$.

\begin{figure}
\begin{center}
\epsfig{file=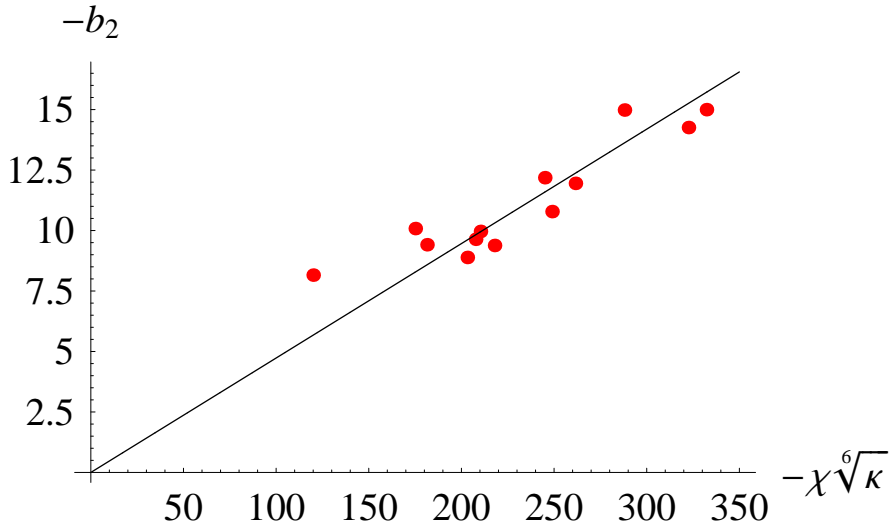, width=12cm}
\end{center}
\medskip
\caption{The plot of $-b_2$ vs. $(-\chi\kappa^{\frac{1}{6}})$ for
13 Calabi-Yau models.} \label{euler-03-17}
\end{figure}

Assuming the constant map contribution is the most significant
contribution at this order in $\CQ$, the coefficient $b_2$ should
behave like
\begin{eqnarray} \label{genustwo-03-17}
b_2 \sim \chi \kappa^{\frac{1}{6}}.
\end{eqnarray}
We can verify the relation (\ref{genustwo-03-17}) by plotting
$b_2$ against the Euler number $\chi\kappa^{\frac{1}{6}}$ for the
13 Calabi-Yau models we studied. We find as the best fit
coefficient
\begin{eqnarray}
b_2=0.047 \chi\kappa^{\frac{1}{6}},
\end{eqnarray}
see the plot in Figure \ref{euler-03-17}, which is reasonably
consistent with the expectation (\ref{genustwo-03-17}).

{}From the second row of the Table \ref{constant-03-29}, we can find
the numerical values of the genus two constant map contribution
$b_2\sim 0.00017g_s^2 \chi\kappa^{\frac{1}{6}}$. Taking into
account that $g_s \sim \mathcal{O}(10)$, this is the same order
of magnitude as our estimate value of $0.047$ from microscopic
topological string computation.

\subsection{Spinning black holes}

We can try to extract the spin dependence of the black hole
entropy from (\ref{entropy1}). Assuming $\CQ \gg J$, and expanding in $J/\CQ$,
we find the following macroscopic
prediction for the topological string data,
\begin{eqnarray} \label{spindependence}
g_m(d)\equiv \frac{d^{\frac{3}{2}}}{m^2}
\log \biggl(\frac{\Omega(d,0)}{\Omega(d,m)}\biggr)=p_0+\mathcal{O}\biggl(\frac{1}{d}\biggr)
\end{eqnarray}
where
\be
\label{asympo}
p_0=3\pi\biggl(\frac{\kappa}{2}\biggr)^{\frac{1}{2}}.
\ee

For a fixed value $m$ we use again the Richardson extrapolation
method to find the asymptotic value of $g_m(d)$ for large $d$. We
list the values of $g_m(d)$ and its first Richardson extrapolation
$A_m(d,1)$ for spin $m=1,2,3$, using the quintic as an example.

We note that the contribution to entropy from angular momentum is
proportional to $d^{-3/2}$, as compared to the
leading static contribution (\ref{entropy2}) of order
$d^{3/2}$. Although the prediction (\ref{spindependence})
should be the leading spinning contribution, there could be some
small statistical fluctuation of topological string data which is
random for the different spins, and which might become comparable
to the spinning contribution in (\ref{spindependence}) and result
in the deviation for large degree $d$. This can be seen in the
quintic example in Table \ref{table5-03-17}. We find that the
Richardson series does not converge to an asymptotic value, instead
the series approach a maximal value before deviating again for large
degree $d$. In order to minimize the effect of statistical
fluctuation of topological string data, we propose to use the
extremal values in the Richardson series $A_m(d,1)$ to estimate
the asymptotic value of $p_0$.
This is indeed a relatively good estimate for the quintic case
where $p_0=14.9019$. Other Calabi-Yau manifolds are analyzed in~\cite{webpage}.

\begin{table}
\begin{centering}
\begin{tabular}{|r|r|r|r|r|r|r|}
\hline   $d$   & $g_1(d)$ & $A_1(d,1)$ & $g_2(d)$ & $A_2(d,1)$ &$g_3(d)$ & $A_3(d,1)$  \\
\hline 1 & 0.693147 & 3.22789 & NA & NA & NA & NA  \\
\hline 2 & 1.96052 & 6.85432 & NA & NA & NA & NA  \\
\hline 3 & 3.59178 & 10.9389 & 9.03347 & 12.2117 & NA & NA  \\
\hline 4 & 5.42856 & 14.4696 & 9.82804 & 13.0403 & 12.1257 & 6.55334  \\
\hline 5 & 7.23677 & 16.4156 & 10.4705 & 12.8183 & 11.0112 & 10.2148  \\
\hline 6 & 8.76658 & 16.1819 & 10.8618 & 11.6135 & 10.8785 & 9.98996  \\
\hline 7 & 9.82591 & 13.9173 & 10.9692 & 9.71239 & 10.7516 & 8.81357  \\
\hline 8 & 10.3373 & 10.4832 & 10.8121 & 7.53259 & 10.5093 & 7.27017  \\
\hline 9 & 10.3535 & 7.02869 & 10.4477 & 5.51774 & 10.1494 & 5.73628  \\
\hline 10 & 10.021 & 4.41912 & 9.9547 & 3.9872 & 9.70809 & 4.46946  \\
\hline 11 & 9.51178 & 2.9195 & 9.4122 & 3.04128 & 9.23185 & 3.58335  \\
\hline 12 & 8.96242 & NA & 8.88129 & NA & 8.76114 & NA  \\
\hline
\end{tabular}  \caption{\label{table5-03-17} The Richardson method for the quintic with spin $m=1,2,3$.}
\end{centering}
\end{table}

We analyze the 13 Calabi-Yau models using the above approach.
Let us define the extremal value of the first Richardson extrapolation
over the degree $d$ as
\begin{eqnarray}
\tilde{g}_m=A_m(d,1)|_{\textrm{max}},
\end{eqnarray}
For various Calabi-Yau models and spin $m=1,2,3$, we compare the
value of $\tilde{g}_m$ with the expected coefficient
$p_0$ given in (\ref{asympo}). The results are
summarized in Table \ref{table6-03-17}. We see that for larger angular
momentum $m$ the deviations become bigger, as expected.

\begin{table}
\begin{centering}
\begin{tabular}{|r|r|r|r|r|r|r|r|}
\hline Calabi-Yau & $p_0=3\pi(\frac{\kappa}{2})^{\frac{1}{2}}$ &
$\tilde{g}_1$ & $\tilde{g}_1$ error & $\tilde{g}_2$ &
$\tilde{g}_2$ error & $\tilde{g}_3$ &
$\tilde{g}_3$ error \\
\hline\ $X_{5}(1)$ & 14.9019 & 16.4156 & 10.2\% & 13.0403 &
-12.5\% & 10.2148 & -31.5\%  \\ \hline   $X_{6}(1^4,2)$ & 11.5429
& 12.1492 & 5.25\% & 10.1828 & -11.8\% & 8.21085 & -28.9\%
 \\
\hline   $X_{8}(1^4,4)$ & 9.42478 & 10.4854 & 11.3\% & 8.1382 &
-13.7\% & 5.3473 & -43.3\%
\\
\hline  $X_{10}(1^3,2,5)$ & 6.66432 & 6.77436 & 1.65\% & 5.89201 &
-11.6\% & 3.62439 & -45.6\%
 \\ \hline  $X_{3, 3}(1^6)$
&19.993 & 22.1786 & 10.9\% & 17.7804 & -11.1\% & 14.8114 & -25.9\%
\\
\hline  $X_{4, 2}(1^6)$ &18.8496 & 21.0741 & 11.8\% & 16.569 &
-12.1\% & 12.9935 & -31.1\%
\\
\hline  $X_{3, 2, 2}(1^7)$ &23.0859 & 25.9065 & 12.2\% & 20.4996 &
-11.2\% & 16.5636 & -28.3\%
\\
\hline  $X_{2, 2, 2, 2}(1^8)$ &26.6573 & 30.1999 & 13.3\% &
23.6923 & -11.1\% & 19.2311 & -27.9\%  \\
\hline  $X_{4, 3}(1^5,2)$ &16.3242 & 17.7685 & 8.85\% & 14.4772
& -11.3\% & 12.2514 & -24.9\%   \\
\hline  $X_{6, 2}(1^5, 3)$ &13.3286 & 15.2332 & 14.3\% &
11.2819 & -15.4\% & 8.06844 & -39.5\%  \\
\hline  $X_{4, 4}(1^4,2^2)$ &13.3286 & 13.9081 & 4.35\% & 11.618 &
-12.8\% & 10.6901 & -19.8\%
\\
\hline  $X_{6, 4}(1^3,2^2, 3)$ &9.42478 & 9.02611 & -4.23\% &
7.87731 & -16.4\% & 7.56862 & -19.7\% \\
\hline $X_{6, 6}({1^2, 2^2, 3^2})$ &6.66432 & 5.42333 & -18.6\%
& 4.91355 & -26.3\% & 4.5984 & -31.0\%  \\
 \hline
\end{tabular}  \caption{\label{table6-03-17} The Richardson method for the 13 Calabi-Yau models with spin $m=1,2,3$.}
\end{centering}
\end{table}

\section{Asymptotics of the Donaldson--Thomas invariants}

As we already mentioned, the  total free energy of the topological string (\ref{totalf}) can be
reorganized in terms of
Gopakumar--Vafa invariants as in (\ref{gova}). A remarkable property of (\ref{gova}) is that for a given class
$Q \in H_2(X,\mathbb{Z})$, the expression is exact in
the string coupling. This is because Castelnuovo's
theorem for the ambient space yields $n_d^g=0$ for $d > \alpha
\sqrt{g}$ for certain $\alpha$.

\begin{figure}[hbtp]
\begin{center}
\epsfig{file=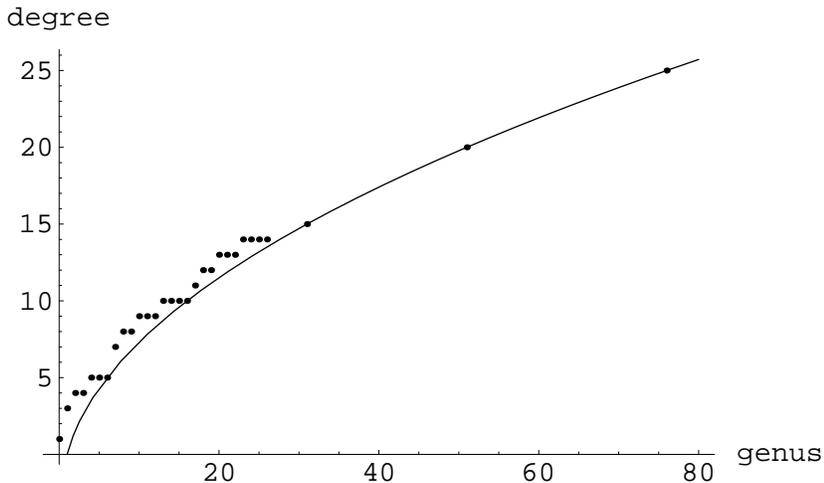, width=11cm}
\end{center}
\caption{Castelnuovo's bound for higher genus curves on the
quintic. The dots represent $n_{d}^{g_{\rm max}}$ and the curve is
(\ref{bound}).} \label{Castelnuovo-03-17}
\end{figure}

For example, for the quintic the maximal genus $g_{\rm max}$ such that
$n_Q^{g_{\rm max}}\neq 0$ fulfills a bound
\begin{equation}
g_{\rm max} \le \frac{1}{10}(10+ 5 d +d^2) \ \label{bound}
\end{equation}
with a decreasing relative deviation in the large $d$ limit. The bound is saturated for curves of total degree $5 m$ which are
complete intersections of degree $(1,5,m)$ in $\mathbb{P}^4$,
which are smooth curves in the quintic. For $5>m>1$ we can
describe the moduli space of the D2 brane as follows. The
linear constraint has as a parameter space $\mathbb{P}^4$  and
allows to eliminate one variable from the degree $m$ constraint,
which has as many homogeneous parameters as monomials in four
variables, i.e. as many as there are integer solutions to
$\sum_{i=1}^4 n_i=m$ namely $\left(m+4-1\atop m\right)$. The
moduli spaces of the curves are therefore fibrations of
$\mathbb{P}^{\left(m+4-1\atop m\right)-1}$ over $\mathbb{P}^4$. Using the
results of \cite{kkv} we get for the GV invariant
\begin{equation}
n_{5 m}^{g_{\rm max}}=(-1)^{\left(m+4-1\atop m\right)-1} \cdot 5 \cdot
\left(m+4-1\atop m\right).
\end{equation}
If the bound (\ref{bound}) is not saturated for small $d$ the
relative deviation can become somewhat larger as seen in the
Figure \ref{Castelnuovo-03-17}.

Let us denote by $F'(\lambda, t)$ the total free energy without the
contribution (\ref{constantmap-03-17}). After exponentiation one finds~\cite{kkrs}
\begin{equation}
Z'_{\rm GV}(X,q,t)=\prod^\infty_{d=1}\left[
\left(\prod_{r=1}^\infty (1-q^r \re^{-dt})^{r n_d^0}\right)
\prod_{g=1}^\infty \prod_{l=0}^{2g-2}(1-q^{ g-l-1}
\re^{-dt})^{(-1)^{g+l} \left(2 g-2\atop l\right) n_d^{g}}\right], \label{zhol}
\end{equation}
where
\be
q=\re^{\ri \lambda}
\ee
and we have assumed that there is only one K\"ahler parameter, so that $Q$ is labeled by a single
integer $d$. On the other hand, the conjecture of \cite{mnop} relating the Donaldson--Thomas invariants $D_{d,n}$
to Gromov--Witten invariants leads to
\begin{equation}
Z_{DT}(q,t)=\sum_{d,n} D_{d,n} q^n \re^{-dt}= Z'_{\rm
GV}(-q,t)M(-q)^{\chi(X)},
\end{equation}
where
\be
M(q)=\prod_{n=1}^\infty\frac{1}{(1-q^n)^n}
\ee
is the MacMahon
function. This  term reinstalls the constant map contribution. We
list for reference a few Donaldson--Thomas invariants $D_{d,n}$ on
the quintic in Table \ref{table7-03-17}.

\begin{table}
\begin{centering}
\begin{tabular}{|r|rrrrrr|}
\hline
d/n & -3 &-2 &-1& 0& 1& 2  \\
\hline
0&  0& 0& 0& 0& 2875& 569250 \\
1&  0& 0& 0& 0& 609250& 124762875 \\
2&  0& 0& 0& 609250& 439056375& 76438831000 \\
 3&  0& 8625& 2294250& 4004590375& 1010473893000& 123236265797125 \\
 \hline
\end{tabular}  \caption{\label{table7-03-17} Donaldson-Thomas invariants.}
\end{centering}
\end{table}

After an extensive discussion of possible tests of the OSV
conjecture \cite{osv}, the  authors of~\cite{dm} isolate as a crucial
question for the validity of the latter the growth behaviour of
the Donaldson-Thomas invariants. This behaviour is encoded in the scaling
exponent $k$, defined as
\begin{equation}
\label{dtscaling}
\log(D_{ \lambda^2 d, \lambda^3 n})\sim \lambda^k\ .
\end{equation}
The question is relevant in the range  $d^3-n^2>0$ for which
stable black hole configurations exist.

Because of Castelnouvo's bound,  and
since our data are up to genus 31, we can calculate the Donaldson--Thomas invariants 
exactly in the range $0\leq d\leq 15$ and for
arbitrary high $n$ for the quintic. We are interested in the limit
\begin{equation}
\label{limit}
k=\lim_{\lambda \rightarrow \infty}\frac{ \log\log| D_{ \lambda^2
d, \lambda^3 n}|}{\log \lambda}\ .
\end{equation}
In order to evaluate it for given values $(d,n)$ we chose
$\lambda$ so that $d+l=\lambda^2 d$ for $d,l\in \mathbb{N}$ and use the fact
that $\log|D_{d+l, n}|$ for fixed $d,k$ scales in good approximation
linearly with $n$ to calculate the interpolated value of the
$D_{d+l,n'}$  at $n'=\lambda^3(d,l) n$, with $\lambda(d,l)=\sqrt{\frac{d+l}{d}}$.
For $(d,0)$  the latter interpolation is of course completely
irrelevant and for charges for which the $n'$ values become
large it is not very relevant.

\begin{figure}
\begin{center}
\epsfig{file=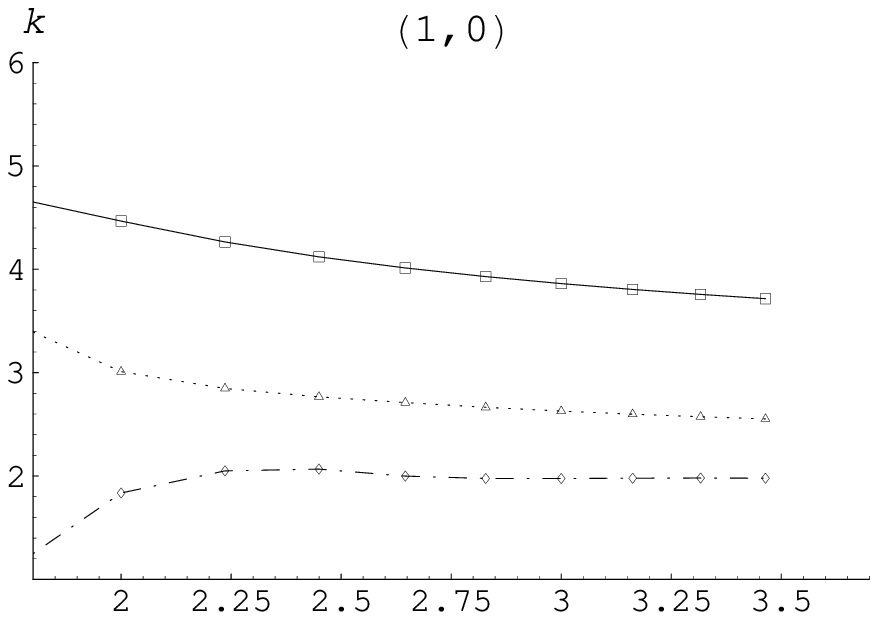,width=8cm}\epsfig{file=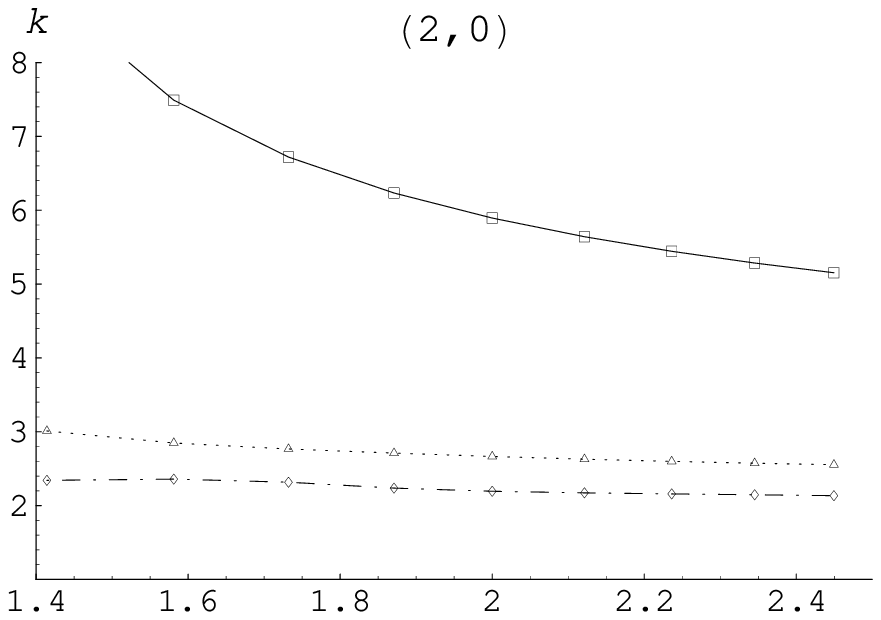,width=8cm}
\end{center}
\vskip -1.2 truecm
{\hskip 7.8 truecm $\lambda$ \hskip 7.4 truecm   $\lambda $ }
\vskip .1truecm
\caption{Scaling data $k^{(0)}$ ({\tiny{$\Box $}}) and the transforms $k^{(1)}$ 
({\tiny{$\triangle$}}), $k^{(2)}$ ($\diamond$) for the Donaldson-Thomas 
invariants on the quintic  in $\mathbb{P}^4$ 
starting for $(d,0)$ states.} \label{DTdx0quintic}
\end{figure}

\begin{figure}
\begin{center}
\epsfig{file=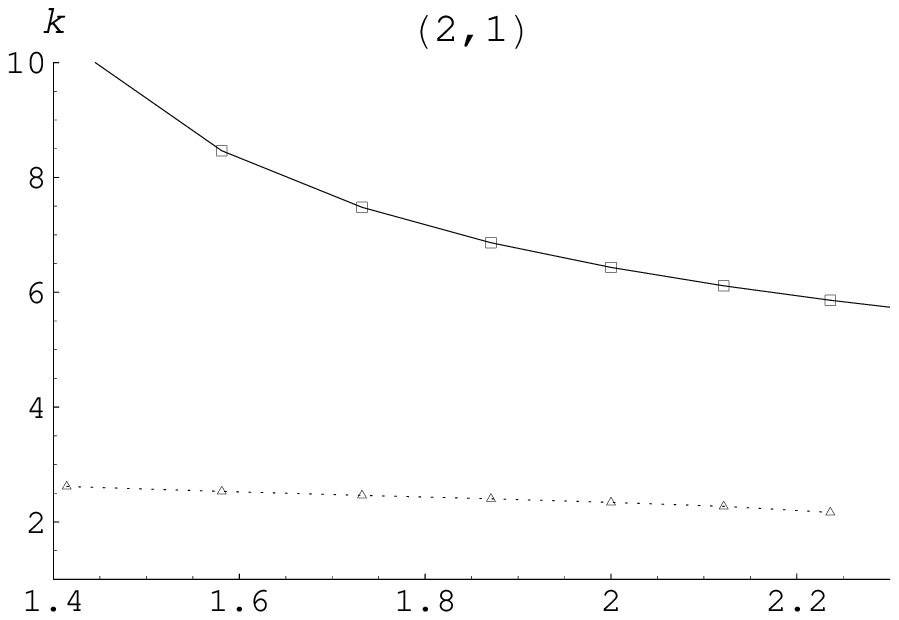,width=8cm}\epsfig{file=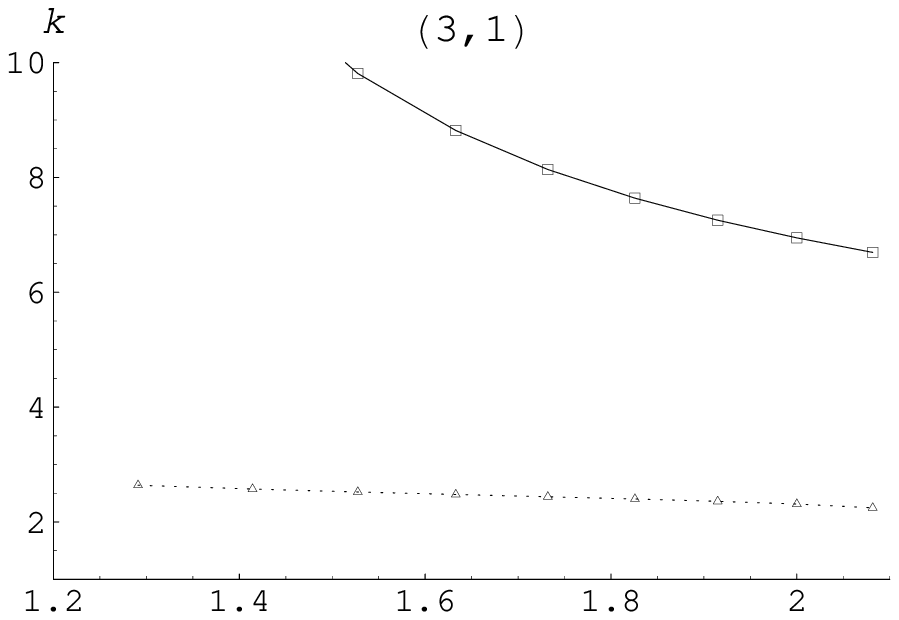,width=8cm}
\end{center}
\vskip -1.2 truecm
{\hskip 7.8 truecm $\lambda$ \hskip 7.4 truecm   $\lambda $ }
\vskip .1truecm
\caption{Scaling data $k^{(0)}$ ({\tiny{$\Box $}}) and the first transform  
$k^{(1)}$ ({\tiny{$\triangle$}})  for the Donaldson-Thomas invariants on 
the quintic in $\mathbb{P}^4$ for the $(2,1)$ and 
$(3,1)$ states.} \label{DTdx1quintic}
\end{figure}

\begin{figure}
\begin{center}
\epsfig{file=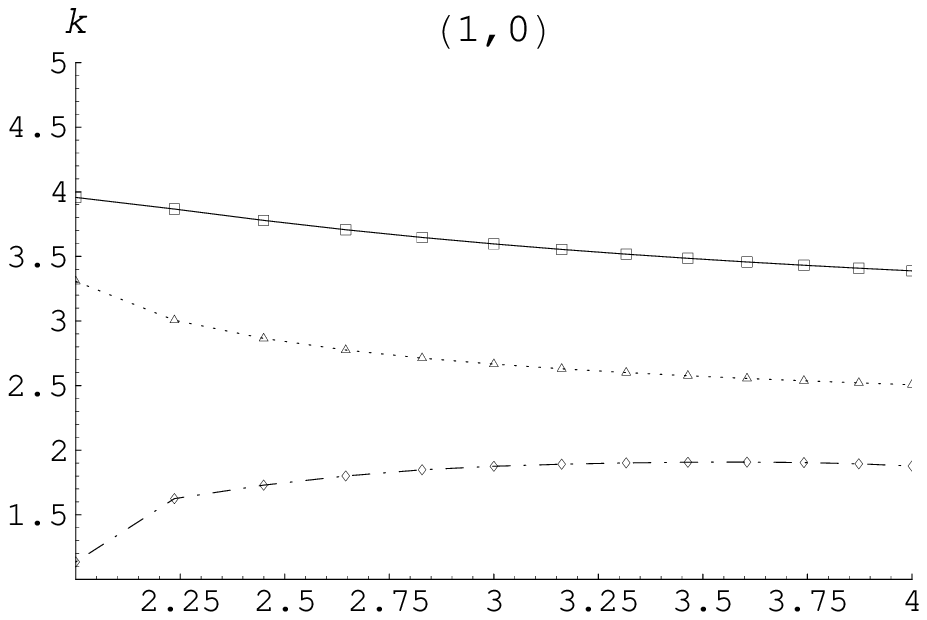,width=8cm}\epsfig{file=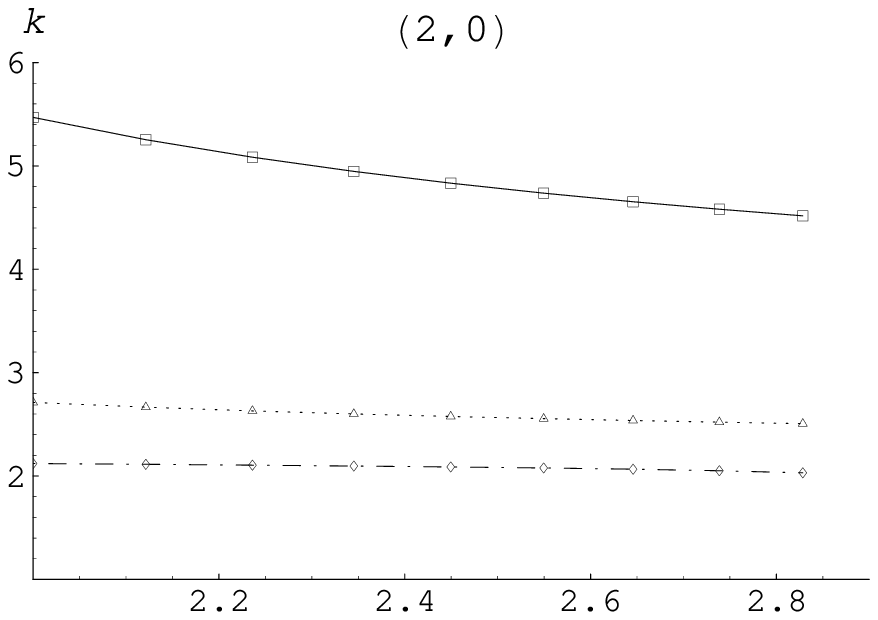,width=8cm}
\end{center}
\vskip -1.2 truecm
{\hskip 7.8 truecm $\lambda$ \hskip 7.4 truecm   $\lambda $ }
\vskip .1truecm
\caption{Scaling data $k^{(0)}$ ({\tiny{$\Box $}}) and the transforms $k^{(1)}$ 
({\tiny{$\triangle$}}), $k^{(2)}$ ($\diamond$) for the Donaldson-Thomas invariants 
on the bic-cubic complete intersection in $\mathbb{P}^5$ 
starting for $(d,0)$ states.} \label{DTdx0bicubic}
\end{figure}

\begin{figure}
\begin{center}
\epsfig{file=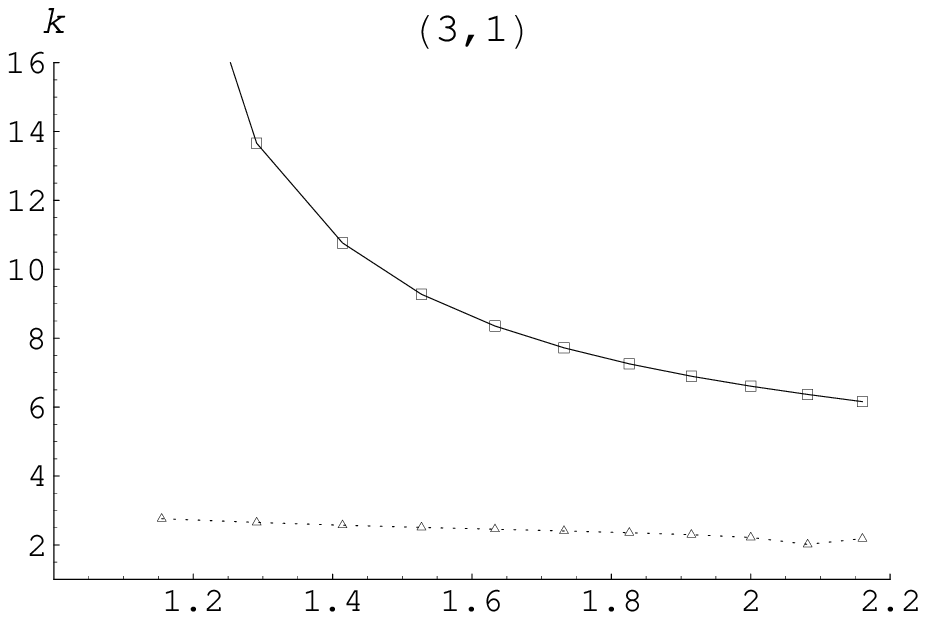,width=8cm}\epsfig{file=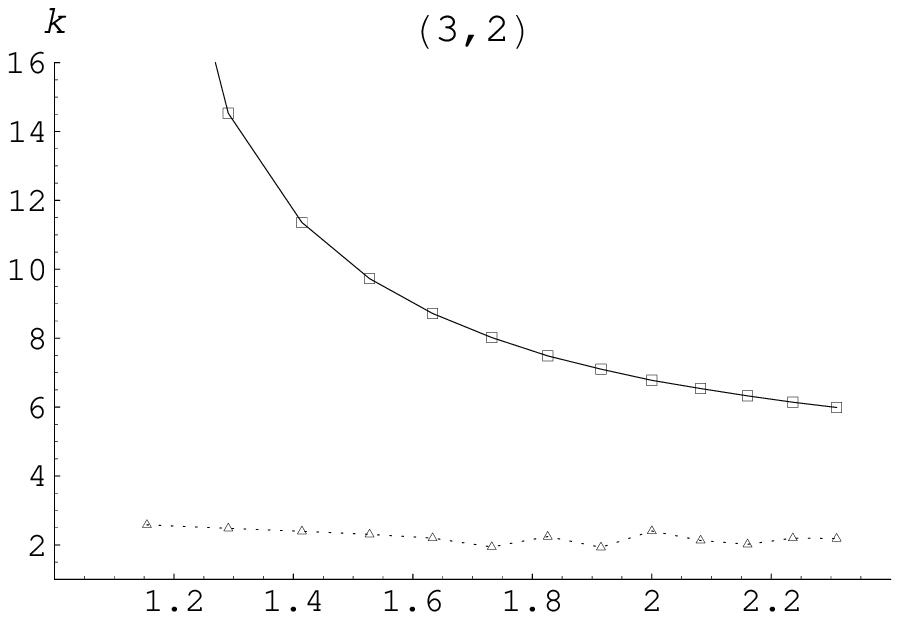,width=8cm}
\end{center}
\vskip -1.2 truecm
{\hskip 7.8 truecm $\lambda$ \hskip 7.4 truecm   $\lambda $ }
\vskip .1truecm
\caption{Scaling data $k^{(0)}$ ({\tiny{$\Box $}}) and the first transform  
$k^{(1)}$ ({\tiny{$\triangle$}}) for the
Donaldson-Thomas invariants on the bic-cubic complete intersection in $\mathbb{P}^5$ for the $(3,1)$ and 
$(3,2)$ states.} \label{DTdx1bicubic}
\end{figure}

The leading correction to (\ref{limit}) is of order
$1/(\log(\lambda))$. It makes therefore sense to eliminate
this leading correction by logarithmic Richardson--Thomas transforms.
We define
\begin{equation}
k^{(0)}_{l}=\frac{\log\log| D_{ \lambda(d,l)^2
d, \lambda(d,l)^3 n}|}{\log \lambda(d,l)},
\end{equation}
and the  $m$th logarithmic Richardson-Thomas transform as
\begin{equation}
k^{(m)}_l=\frac{ k^{(m-1)}_{l+1}\log(l+1)- k^{(m-1)}_{l}\log(l)}{\log(l+1)-\log(l)}.
\end{equation}
With our knowledge of the topological string up to $g=31$ for the quintic we can evaluate
the Donaldson-Thomas invariants up to degree $15$.  We plot in the first two graphs
(\ref{DTdx0quintic}) the data for the $k^{(0)}_l$ and its  first two logarithmic
Richardson-Thomas transforms.  The graphs clearly indicate
that the convergence is improved by the transform. So even if there
are subleading terms of other forms, we certainly managed to supress the
leading correction and speed up the convergence. The data further show
that there is an universal behaviour independent of $d$ and that the
value of $k$ is within the $2\%$ range close to $2$. The higher
logarithmic Richardson-Thomas transforms are consistent with
this value but do not determine it better as we also have
to take into account values with smaller $l$ hence smaller
$\lambda$. We next test the universality of these results
for other charges $(d,n)$ in \figref{DTdx1quintic}. If $n\neq0$ we need the
interpolation for the $n'$ values. This introduces some
random subleading errors, which are of the order of the improvement
by the second logarithmic Richardson-Thomas transform. However as
in the figure for $(2,0)$ we see that higher $d$ seems to lower
the coefficient of the  sub-sub-leading  correction and makes already the second
Richardson transform to converge reasonably well --well enough at least to
conclude that the $k$ is considerably lower then $3$ and very well compatible
with the value $k=2\pm 0.03$ found for the previous charges. We solved 
the bicubic in $\mathbb{P}^5$ up to genus $29$, which yields complete 
informations about the Donaldson-Thomas invariants up to degree $18$. A similar 
analysis as above confirms the analysis for the quintic. The corresponding plots 
are in Fig. \ref{DTdx0bicubic} and  Fig. \ref{DTdx1bicubic}. Again a detailed
summary of the data for more models can be found at~\cite{webpage}. 
We note a slight noise in the  transform $k^{(1)}$ in Fig. \ref{DTdx1bicubic}, 
which is presumably due to the interpolation in the $n$ value of $D_{d,n}$ 
described above. The results for the other models are similar, but somewhat 
less precise due to smaller values of $d$ that are currently available.

To summarize: our analysis indicates that the value of $k$ is indeed universal  and close to $k=2$. This
strongly suggests that the ``mysterious cancellations" \cite{dm} that eventually
make possible to extend the the OSV conjecture to small coupling, actually take place.

\sectiono{K3 fibrations}

\subsection{Topological strings on K3 fibrations}

We will now consider Calabi--Yau manifolds $X$ that have the structure of a K3 fibration, i.e. there is a fibration of the form
\be
\pi: X \rightarrow \IP^1,
\ee
where the fibers are K3 surfaces. When the fibration is regular the homology of $X$ can be written as
\be
H_2(X,\IZ) =\langle [\IP^1]\rangle \oplus {\rm Pic}({\rm K3}),
\ee
where ${\rm Pic}({\rm K3})$ is the Picard lattice of the K3 fiber. The rank of this lattice will be denoted by
$\rho$, and $\Sigma^a$, $a=1, \cdots, \rho$ will denote a basis for this
lattice. Let $\omega$ be the complexified K\"ahler form on $X$. The complexified K\"ahler parameters of $X$ are given by
\be
S=\int_{\IP^1} \omega, \qquad t^a =\int_{\Sigma^a} \omega, \quad a=1, \cdots, \rho.
\ee
We will denote by $\eta_S$, $\eta_a$ the two--forms which are dual
to $\IP^1$, $\Sigma^a$.

It turns out that type IIA string theory compactified on these manifolds is very often
dual to heterotic string theory compactified on K3$\times \IT^2$ \cite{kv,klm}. Under this duality,
$S$ becomes the axidilaton of the heterotic string. It follows
that in the regime $S\rightarrow \infty$ one can map computations in the type IIA theory to perturbative computations in the heterotic string. In particular,
the $F_g$ couplings of topological string theory (which are graviphoton couplings in type IIA theory) can be computed exactly at one--loop in the heterotic string, provided
the K\"ahler parameters are restricted to the K3 fiber \cite{kv,klm,hm,agnt}. We will now review here some of these results.

The topological string amplitudes $F_g(S,t)$ on these fibrations have the following structure,
\be
\label{fgfibs}
\ba
F_0(S,t)& ={1\over 6}C_{abc} t^a t^b t^c +{1\over 2} C_{ab}S t^a t^b +{\zeta (3)\over 2} \chi(X) +\CF_0(t) +\CO(\re^{-S}), \\
F_1(S,t)&={1\over 24}(c_S S + c_at^a) +\CF_1(t) +\CO(\re^{-S}),\\
F_g(S,t)&=d_g \chi(X) + \CF_g(t) + \CO(\re^{-S}), \qquad g\ge 2.
\ea
\ee
In these formulae, $C_{abc}$ and $C_{ab}$ are triple intersection numbers in the fiber and in the mixed fiber/base direction, respectively.
Notice that
\be
C_{abc}=\int_X \eta_a \wedge \eta_b \wedge \eta_c, \qquad C_{ab}=\int_X \eta_S \wedge \eta_a \wedge \eta_b.
\ee
We also have
\be
c_a =\int_X c_2(X) \wedge \eta_a,  \quad a=1, \cdots, \rho, \qquad c_S=\int c_2(X) \wedge \eta_S.
\ee
For K3 fibrations with trivial fundamental group one has $c_S=24$~\cite{Oguiso}, but for the Enriques Calabi--Yau 
(which we will also analyze), $c_S=12$. The coefficient $d_g$ is the contribution
of constant maps written down in (\ref{dg}).
In (\ref{fgfibs}), $\CF_g(t)$ denotes the contribution of
worldsheet instantons in the K3 fiber. It follows from \cite{agnt,hm,kawai,mm,kkrs} that
the $\CF_g(t)$ can be completely determined in terms of a single modular form that
we will denote $f_X(q)$. In order to write down an explicit formula for $\CF_g(t)$, we have to introduce the quasimodular forms $\CP_g(q)$ which are defined by
\be
\label{defpg}
\biggl( { 2\pi  \eta^3 \lambda \over \vartheta_1(\lambda|\tau)}\biggr)^2=
\sum_{g=0}^{\infty} (2 \pi \lambda)^{2g} {\cal P}_{g}(q).
\ee
The quantities ${\cal P}_{g}(q)$ can be explicitly written in terms of generalized Eisenstein series \cite{mm}, and one has for example
\be
\label{casesps}
\CP_1(q)={1\over 12}E_2(q), \,\,\,\,\,\, \CP_2(q)={1 \over 1440} (5 E_2^2 + E_4).
\ee
We now introduce the coefficients $c_g(n)$ through
\be
\label{defcg}
{\cal P}_{g} (q) f_X(q)=\sum_n c^{X}_g(n) q^n.
\ee
One then has the following expression for the heterotic $\CF_g(t)$:
\be
\label{fgex}
\CF_g(t)= \sum_{Q\in {\rm Pic}({\rm K3})} c^{X}_g(Q^2/2){\rm Li}_{3-2g}({\rm e}^{-Q\cdot t}),
\ee
where ${\rm Li}_{n}$ is the polylogarithm of index $n$
\be
{\rm Li}_n (x) =\sum_{k=1}^{\infty} {x^k \over k^n}.
\ee
In (\ref{fgex}) we have also denoted
\be
Q\cdot t = n_a t^a, \qquad Q^2=C^{ab}n_a n_b,
\ee
where $C^{ab}=C_{ab}^{-1}$ is the intersection form of the Picard lattice ${\rm Pic}({\rm K3})$.

We will particularly interested in three special K3 fibrations: the STU model, the ST model, and the Enriques Calabi--Yau. Let us give some extra details for these cases:

\begin{itemize}

\item The STU model has $\rho=2$ and it can be realized by a complete intersection in a weighted projective space which
is frequently denoted by $X_{24}(1,1,2,8,12)$. It has Euler characteristic $\chi=-480$. The classical prepotential can be
obtained from the nonvanishing intersection numbers,
\be
C_{111} =8, \quad C_{112}=2, \quad C_{11}=2, \quad C_{12} =1,
\ee
while the classical part of $F_1(S, t)$ is encoded by
\be
c_1=92, \quad c_2=c_S=24.
\ee
The modular form encoding the information about topological string amplitudes in the fiber is given by \cite{mm}
\be\label{STUform}
f_{\rm STU}(q)=  -{2 E_4 E_6  \over \eta^{24}}(q).
\ee
It is sometimes useful to parametrize the K\"ahler cone in terms of the variables
\be
T=t_1+ t_2, \qquad U=t_1,
\ee
In this basis one has $Q^2/2=mn$.

\item The ST model has $\rho=1$ and is realized in type IIA by the CY $X_{12}(1,1,2,2,6)$. It has $\chi=-252$ and the
classical intersection numbers
\be
C_{111}=4, \quad C_{11}=2,
\ee
as well as
\be
c_1=52,\qquad c_S=24.
\ee
The K\"ahler parameter along the fiber is usually denoted as
\be
T=t_1.
\ee
The relevant modular form is \cite{kawai,kkrs}
\be
\label{STform}
f_{\rm ST}(q)=-{2 \theta E_4 F_6  \over \eta^{24}}(q),
\ee
where
\be
\ba
\theta(q)& =\sum_{n \in \IZ} q^{n^2 \over 4}=\vartheta_3(\tau/2), \\ F_2&={1\over 16} \vartheta_2^4(\tau/2), \\F_6& =E_6-2 F_2 (\theta^4 -2 F_2)(\theta^4 -16 F_2).\ea
\ee
Notice that $Q^2/2=n^2/4$.

\item The Enriques Calabi--Yau is given by the free quotient $({\rm K3} \times \IT^2)/\IZ_2$, and was introduced in the context of
type II/heterotic duality in \cite{fhsv}. It is an elliptic fibration with $\rho=10$. It has $C_{abc}=0$, while $C_{ab}$ is given by the intersection
numbers of the Enriques surface $E$. The Picard lattice is
\be
{\rm Pic}({\rm K3})=\Gamma^{1,1} \oplus E_8(-1),
\ee
and
\be
c_a=0, \qquad c_S=12.
\ee
The topological string amplitudes in the fiber were obtained in \cite{km} (see also \cite{gkmw}). They are also controlled by a single
modular form
\be\label{Eform}
f_E(q)=-{2\over \eta^{12}(q^2)},
\ee
but their form is slightly different from (\ref{fgex})
\be
F_g(t)=\sum_{Q\in  {\rm Pic}({\rm K3})} c^E_g(Q^2) \bigg\{ 2^{3-2g} {\rm Li}_{3-2g}(\re^{-Q\cdot t}) - {\rm Li}_{3-2g}(\re^{-2 Q\cdot t})\biggr\},
\label{heteroticprediction}
\ee
where $c^E_g(n)$ are defined again by (\ref{defcg}).
\end{itemize}

\subsection{Microscopic degeneracies and their asymptotic expansion}

We have seen that, at least in the case of topological strings on K3 fibrations, and for classes $Q$ restricted to the K3 fiber, one can obtain
closed formula for the topological string amplitudes at all genera. It should be therefore possible to extract a closed
formula for the generating functional of Gopakumar--Vafa invariants. In fact, by using the product formula
\be
\label{prodone}
\vartheta_1(\nu|\tau)=-2 q^{1\over 8} \sin (\pi \nu) \prod_{n=1}^{\infty} (1-q^n) (1-2 \cos (2 \pi \nu) q^n + q^{2n})
\ee
one finds from the expression (\ref{fgex}) and the structure (\ref{gova})
\be
\label{gvhet}
\sum_{Q \in {\rm Pic (K3)}}\sum_{r=0}^{\infty}
n^r_Q z^{2r} p^{Q^2/2} =f_X (p) \xi^2(z),
\ee
where $\xi(z)$ is the function that appears in helicity supertraces,
\be
\label{helicityxi}
\xi(z)=\prod_{n=1}^{\infty} {(1-p^n)^2 \over (1-p^n)^2 + z^2 p^n}=\prod_{n=1}^{\infty} {(1-p^n)^2 \over (1-p^n y)(1-p^ny^{-1})},
\ee
where we have set $z=-\ri(y^{1\over 2} -y^{-{1\over 2}})$.

We can now obtain a closed formula for the microscopic degeneracies. In order to have a description
which incorporates as well the elliptic genus, we will count the microstates as in (\ref{degs}) but with
$r\rightarrow r-1$. With this definition, the l.h.s. of (\ref{gvhet}), expanded in $q,y$,
is precisely the generating function of microscopic degeneracies $\Omega(Q,m)$, summed over
all $m, Q$. We then arrive to the expression
\be
\label{exdegs} \sum_{Q\in {\rm Pic}({\rm K3})} \sum_{m=-\infty}^{\infty}\Omega(Q,m)y^mp^{Q^2/2}=f_X(p)\xi^2(\nu,\sigma),
\ee
where we have written
\be
y=\re^{2 \pi\ri \nu}, \qquad p=\re^{2\pi \ri \sigma}.
\ee
Notice that if we consider $X={\rm K3} \times \IT^2$ and restrict to classes $Q$ in the fiber, the counting
of microstates given by the elliptic genus is
\be
\chi(S_p{\rm K3};q,y)_{q^0}= \prod_{N=1}^{\infty} {1\over (1-p^N)^{20} (1-p^N y)^2 (1-p^Ny^{-1})^2} =
{p\over \eta^{24}(p)} \xi^2(y).
\ee
This has the same form than (\ref{exdegs}) with
\be
f_{{\rm K3} \times T^2}(p)=\frac{1}{\eta^{24}(p)},
\ee
therefore we can consider the ``small" D1--D5 system as a particular case of our analysis.

The expression (\ref{exdegs}) tells us that the microscopic degeneracies we are looking for are simply the Fourier coefficients of the object in the
r.hs. We can then invert it to write
\be
\label{omegaint}
\Omega(N,m)=\int_{-\frac{1}{2}+\ri 0^{+}}^{\frac{1}{2}+\ri  0^{+}}\rd \sigma\int^{1}_{0} \rd\nu \,  \re^{-2 \pi \ri (N\sigma+m\nu)}
\Phi(\nu,\sigma), \qquad N=Q^2/2,
\ee
where we defined
\be
\label{phif}
\Phi(\nu,\sigma)=f_X(p)\xi^2(\nu,\sigma). \ee
and we have assumed that $N$ is a non-negative integer (this can be guaranteed by rescaling $p \rightarrow p^k$ for some appropriate $k$).
The contour in (\ref{omegaint}) has been chosen to avoid the poles in the integrand.

We will now evalute the asymptotic expansion of $\Omega(N) \equiv
\Omega(N,0)$ in inverse powers of $N$. Nonzero values of the spin $m=0$ can be analyzed in a similar way. The expression we will find is exact up to corrections which are exponentially suppressed in the large charge
limit $N \rightarrow \infty$. Notice that in our situation we can not appeal to the Rademacher expansion which was used in \cite{dmmv,ddmp}, since
(\ref{phif}) is not a Jacobi form (it can be regarded as a Jacobi form with {\it negative} index).
It is likely that an analog of the Rademacher expansion exists, but we will perform a direct evaluation of
the integral (\ref{omegaint}) in the spirit of the counting of states with spin in Appendix C of \cite{ddmp} and in \cite{dabholkar}.

First of all, we reexpress the integrand (\ref{phif}) in terms of $\vartheta_{1}(\nu|\sigma)$ as,
\be
\Phi(\nu,\sigma)=4\sin^{2}(\pi \nu)\eta^{6}(p)\frac{f_X(p)}{\vartheta_{1}^2(\nu|\sigma)}.
\ee
Using the modular behavior of $\vartheta_{1}(\nu|\sigma)$ under the $S$ transformation
$\sigma \rightarrow \tilde \sigma =-1/\sigma$ we get,
\be
\label{inverse}
\vartheta_{1}(\nu,\sigma)=-\frac{2 \ri}{\sqrt{-\ri \sigma}}\re^{\frac{\pi}{\ri \sigma}(\nu^2+\frac{1}{4})}\sin\Bigl(\frac{\pi \nu}{\sigma}\Bigr)\Bigl\{1+
O(\re^{-\frac{2\pi\ri}{\sigma}})\Bigr\}.
\ee
It is easy to see that the saddle point evaluation of (\ref{omegaint}) is governed by
\be
\label{saddle}
\sigma_{*}=\frac{\ri}{\sqrt{N}}+\CO\Bigl(\frac{1}{N}\Bigr).
\ee
Therefore, the corrections to (\ref{inverse}) will be exponentially suppressed. Using the modularity of $\eta(p)$, and taking the part of the
$\sin$ in the denominator which is not exponentially suppressed, we obtain,
\be \Phi(\nu,\sigma)\sim -4 \sigma^{-2}\re^{2\frac{\ri \pi}{\sigma}(\nu^2-\nu)}\sin^{2}(\pi \nu) f_X(p).
\ee
Therefore, in order to compute the asymptotics of (\ref{omegaint}) we just need
\be
\Omega(N)\sim -4\int_{-\frac{1}{2}+ \ri 0^{+}}^{\frac{1}{2}+ \ri 0^{+}}\rd\sigma \, \re^{-2\pi \ri N \sigma}\frac{f_X(p)}{\sigma^{2}}\int^{1}_{0}
 \rd\nu\, \re^{2\frac{\ri \pi}{\sigma}(\nu^2-\nu)}\sin^{2}(\pi \nu).
 \ee
The integral over $\nu$ is easily worked out in terms of the error function ${\rm Erf}(x)$, as follows,
\be\label{Erfint}
\ba
&  \int^{1}_{0}
 \rd\nu\, \re^{2\frac{\ri \pi}{\sigma}(\nu^2-\nu)}\sin^{2}(\pi \nu) =
  \sqrt{\frac{\ri \sigma}{8}}\re^{ \frac{\pi}{2 \ri \sigma}}
{\rm Erf}\Bigl( \sqrt{\frac{\pi}{2 \ri \sigma}}\Bigr)\\
&\, \,  + \sqrt{\frac{\ri \sigma}{32}}\re^{ \frac{ \pi}{2 \ri}(\sigma+\frac{1}{\sigma})} \biggl\{{\rm Erf}\Bigl( \sqrt{\frac{\pi}{2 \ri \sigma}}(\sigma+1)\Bigr)-
{\rm Erf}\Bigl(
\sqrt{\frac{\pi}{2 \ri \sigma}}(\sigma-1)\Bigr)\biggr\}.
\ea
\ee
Due to (\ref{saddle}) we can use the asymptotic expansion of the
${\rm Erf}$ function,
\be\label{asympErf}
{\rm Erf}(x) \sim 1-\frac{\re^{-x^2}}{\sqrt{\pi} }\sum_{r=0}^{\infty}(-1)^r \frac{(2r-1)!!}{2^r} x^{-(2 r+1)}, \qquad |x|\rightarrow \infty,\,\, |{\rm arg}(-x)|<\pi.
\ee
Ignoring terms which are exponentially suppressed at large $N$, we find,
\be \int^{1}_{0}
 \rd\nu\, \re^{2\frac{\ri \pi}{\sigma}(\nu^2-\nu)}\sin^{2}(\pi \nu)\sim -\frac{1}{4}\sum_{r=0}^{\infty} \frac{\ri^{1+3 r}}{\pi^{1+r}}(2 r -1)!!
G_{r}(\sigma),
\ee
with,
\be
G_{r}(\sigma)=\sigma^{r+1}\biggl( 2 + \frac{1}{(\sigma-1)^{1+2 r}} - \frac{1}{(\sigma+1)^{1+2 r}}\biggr).
\ee
Again, due to (\ref{saddle}), we can expand it around $\sigma=0$,
\be G_{r}(\sigma)=-2\sum_{s=0}^{\infty}{ 2 (1+s+r) \choose 2 r} \sigma^{3 + 2 s + r}.
\ee
Putting all together, we obtain,
\be
\Omega (N)\sim 2\sum_{r=0}^{\infty}\frac{(2 r-1)!!}{(\ri \pi)^{r+1}}\sum_{s=0}^{\infty} { 2 (1+s+r) \choose 2 r}
\int_{-\frac{1}{2}+ \ri 0^{+}}^{\frac{1}{2}+ \ri 0^{+}}\rd \sigma \, \re^{-2\pi \ri N \sigma}f_X(p)\sigma^{1+2s+r}.
\ee
We now work out the integral,
\be
A_{s,r}(N)\equiv\int_{-\frac{1}{2}+ \ri 0^{+}}^{\frac{1}{2}+ \ri 0^{+}}\rd\sigma \,
\re^{-2\pi \ri N \sigma}f_X(p)\sigma^{1+2s+r}.
\ee
We assume that $f_X(p)$ has modular weight $w$, so that $f_X(p)=
\sigma^{-w}f_X(\tilde{p})$, where $\tilde{p} =\re^{-\frac{2 \pi \ri}{\sigma}}$. For the modular forms that we consider here,
$f_X(\tilde{p})=c \tilde{p}^{-\alpha}+\cdots$, and the integral above gives a modified Bessel function
\be
A_{s,r}(N)\sim c\ri^{1+2 s+r-w}\hat{I}_{2 s+r+2-w}(4\pi\sqrt{\alpha N}). \ee
We end up then with the following result for the exact asymptotics of the microscopic black hole degeneracy,
\be\label{takelog}
\Omega (N) \sim 2c\ri^{w}\sum_{r=0}^{\infty}\frac{(2 r-1)!!}{\pi^{r+1}}\sum_{s=0}^{\infty}(-1)^{s} {2 (1+s+r) \choose 2 r}
\hat{I}_{2 s+r+2-w}(4 \pi \sqrt{\alpha N}).
\ee
Using now the formula for the asymptotic expansion of $\hat{I}$ functions (see for
example App. A of \cite{ddmp}), we find for the entropy $S(N) =\log\, \Omega(N)$ the following 
expansion
\be\label{f5}
S \sim 4 \pi \sqrt{\alpha N} - \frac{5-2 w}{4}\log(N)+\log\biggl(\frac{\sqrt{2} \ri^w \alpha^{\frac{2w-5}{4}}c}{\pi}\biggr)
+\frac{177+16 w-4 w^2}{32 \pi \sqrt{\alpha}}\frac{1}{\sqrt{N}}
+ \CO(N^{-1}).
\ee
The expansion in powers of $1/N^{1\over 2}$ in (\ref{takelog}), which is obtained by using the asymptotics 
of modified Bessel functions, is the expansion of the original integral around the saddle point (\ref{saddle}). This can be 
verified by an explicit computation of the first few orders of the saddlepoint expansion. 
 
Let us now evaluate (\ref{f5}) in some examples. For K3$\times \IT^2$ we have $(w,\alpha, c)  = (-12, 1,1)$, and the entropy reads
\be
S \sim  4 \pi \sqrt{ N} - \frac{29}{4}\log(N)+\log\biggl(\frac{\sqrt{2}}{\pi}\biggr)-\frac{591}{32 \pi}\frac{1}{\sqrt{N}}
+ \CO(N^{-1}).
\ee
For the STU model, with the values $(w,\alpha,c)=(-2,1,-2)$, we find
\be\label{f6}
S \sim 4 \pi \sqrt{ N} - \frac{9}{4}\log(N)+\log\biggl(\frac{\sqrt{8}}{\pi}\biggr)+\frac{129}{32 \pi}\frac{1}{\sqrt{N}}
+ \CO(N^{-1}).
\ee

The ST model is slightly different, since in $f_{\rm ST}(p)$
both integer and rational powers of $p$ appear. As mentioned above, we should redefine $p \rightarrow p^4$ and write down the
generating functional for the degeneracies as
\be
\label{4exdegs} \sum_{Q\in {\rm Pic}({\rm K3})} \sum_{m=-\infty}^{\infty}\Omega(Q,m)y^mp^{2 Q^2}=f_{\rm ST}(p^4)\xi^2(\nu,4 \sigma),
\ee
where we recall that $M \equiv 2 Q^2 = n^2$ is an integer. The asymptotics is given by the integral
\be
\label{omegaintST}
\Omega_{\rm ST}(M) \sim -\int_{-\frac{1}{2}+\ri 0^{+}}^{\frac{1}{2}+\ri  0^{+}}\rd \sigma\;\re^{-2 \pi \ri M\sigma} \frac{f_{\rm ST}(p^4)}{4 \sigma^2}
\int^{1}_{0} \rd\nu \,\sin^2(\pi \nu ) \re^{\frac{\ri \pi}{2 \sigma}(\nu^2-\nu)}.
\ee
The integral over $\nu$ is given by (\ref{Erfint}) upon replacing $\sigma \rightarrow 4 \sigma$. Since,
\be
f_{\rm ST}(p^4)=-2\frac{E_4(p^4)F_6(p^4)}{\eta^{24}(p^4)}\vartheta_{3} (2 \sigma) \sim -16 \sqrt{2 \ri} \sigma^{\frac{3}{2}} \re^{\frac{\ri \pi}{2 \sigma}},
\ee
one finds in the end,
\be
\Omega_{\rm ST}(M) \sim \sqrt{2}\sum_{r=0}^{\infty}\frac{(2 r-1)!!}{\pi^{r+1}}\sum_{s=0}^{\infty}(-1)^{s} {2 (1+s+r) \choose 2 r}
\hat{I}_{\frac{7}{2}+2 s+r}(2 \pi \sqrt{M}),
\ee
and from here one can read the entropy,
\be\label{STentropy}
S(Q)\sim 4 \pi {\sqrt{{1\over 2} Q^2}}-2 \log(Q^2)+\cdots
\ee

Finally we turn to the case of Enriques CY manifold. It follows from (\ref{heteroticprediction}) that one has to distinguish two types of
homology classes: the classes $Q$ whose entries contain at least an odd
integer (which were called odd classes in \cite{km}), and the classes $Q$ for which all entries are even (called even classes).
A simple calculation shows that the generating function of Gopakumar--Vafa invariants for the
odd classes is given by
\be
\label{odd}
\sum_{r=0}^{\infty} \sum_{Q\,\ {\rm odd}} n_Q^r p^{Q^2} z^{r-1}={f_E(q) \over 4 \sin^2 \, \bigl({\pi \nu  \over 2} \bigr)} \Bigl( \xi^2(\nu/2,p) -\xi^2(\nu/2,-p) \Bigr).
\ee
while for the even classes is given by
\be
\label{even}
\ba
\sum_{r=0}^{\infty} \sum_{Q\,\, {\rm even}} n_Q^r p^{Q^2} z^{r-1}&={f_E(q) \over 4 \sin^2 \, \bigl({\pi \nu  \over 2} \bigr)} \Bigl(\xi^2(\nu/2,p) -\xi^2(\nu/2,-p) \Bigr)\\
&  -f_E(q^4)
\Bigl( \xi^2(\nu, p^4) -\xi^2(\nu, -p^4) \Bigr).
\ea
\ee
Notice that for even classes $Q^2\equiv 0$ mod $4$, while for odd classes one only has $Q^2 \equiv 0$ mod $2$. In contrast to the previous K3 fibrations, in the above 
generating function we have $p^{Q^2}$, instead of $p^{Q^2/2}$, and this will lead to a different leading term as compared for example to the STU model.

The computation of the asymptotics of the microstates is similar to the one that we just performed. Let us begin with odd classes. Using the identity,
\be
\xi^{2}(\nu,-p)=4 \sin^{2}(\pi \nu) \frac{ \eta^{6}(2 \sigma) \vartheta_{3}^{2}(2 \sigma)}{\vartheta_{1}^{2}(\nu|2 \sigma)\vartheta_{3}^{2}(\nu|2 \sigma)},
\ee
and proceeding as in the previous case, we find,
\be
 \Omega_{\rm odd}(N)=\Omega_{1}(N) + \Omega_{2}(N),\qquad N=Q^2/2,
 \ee
where,
\be
\ba
\Omega_1(N)&\sim 16 \int_{-\frac{1}{2}+ \ri 0^{+}}^{\frac{1}{2}+ \ri 0^{+}} \rd \sigma \, \re^{-4\pi \ri N \sigma} \sigma^2 \eta^{6}(2 \sigma)
\vartheta_{3}^{2}(2 \sigma) f_E(p) \int_{0}^{1} \rd\nu \sin^2(\pi \nu) \re^{\frac{\ri \pi}{2 \sigma}(\nu^2-\nu+\frac{1}{2})}, \\
\Omega_2(N)&\sim -4 \ri \int_{-\frac{1}{2}+ \ri 0^{+}}^{\frac{1}{2}+ \ri 0^{+}} \rd\sigma \,  \re^{-4\pi \ri N \sigma} \sigma \eta^{6}(\sigma)
f_E(p) \int_{0}^{1} \rd\nu \sin^2(\pi \nu) \re^{\frac{\ri \pi}{2 \sigma}(\nu-1)^2}.
\ea
\ee
As before, we evaluate the integrals over $\nu$ in terms of the Erf function and its asymptotic expansion. We then use the modularity properties of the
different functions involved here to obtain,
\be\label{oddOmega}
\Omega_{\rm odd} (N)\sim \frac{1}{16}\sum_{r=0}^{\infty}\frac{(2 r-1)!!}{\pi^{r+1}}\sum_{s=0}^{\infty} (-1)^s {2 (1+s+r) \choose 2 r}
(1-4^{-(1+r+s)})\hat{I}_{8+2 s+r}(\pi \sqrt{8N}).
\ee
Let us now consider the even classes, (\ref{even}). Comparing (\ref{even}) with (\ref{odd}), we see that,
\be
\Omega_{\rm even}(N)=\Omega_{\rm odd}(N)-\widetilde{\Omega}(N)
\ee
where,
\be\label{ref0}
\tilde{\Omega}(N)= \int_{-\frac{1}{2}+ \ri 0^{+}}^{\frac{1}{2}+ \ri 0^{+}} \rd \sigma \int_{0}^{1} \rd\nu \, \re^{-4 \ri \pi N \sigma}
4 \sin^2(\pi \nu)f_E(p^4)\Bigl( \xi^2(\nu, p^4) -\xi^2(\nu, -p^4) \Bigr).
\ee
A computation similar to the one we performed shows that $\widetilde \Omega (N)$ is exponentially suppressed with respect to
$\Omega_{\rm odd}(N)$, since it leads to terms that go like $\exp(\pi \sqrt{2 N})$ and $\exp(\pi \sqrt{6 N})$. Therefore, as an
asymptotic expansion in $1/{\sqrt {N}}$, $\Omega_{\rm even}(N)\sim \Omega_{\rm odd}(N)$, and the asymptotics does not
distinguish between the even and the odd classes. We finally obtain, for the small Enriques black hole,
\be\label{Enrientropy}
S_{E}(Q)\sim2 \pi {\sqrt {Q^2}} - \frac{17}{2} \log {\sqrt {Q^2}} + \cdots.
\ee

The main conclusion of our analysis is that, in all cases, the leading term of the microscopic entropy for these black holes is given by
\be
\label{microentropy}
S(Q) \sim 2 \pi {\sqrt {  {c_S \over 12} Q^2}},
\ee
since $c_S=24$ for K3$\times \IT^2$, the STU and the ST models, but $c_S=12$ for the Enriques CY. Of course, our analysis has also
given precise formulae for the subleading terms. 

\begin{figure}[hbtp]
\begin{center}
\epsfig{file=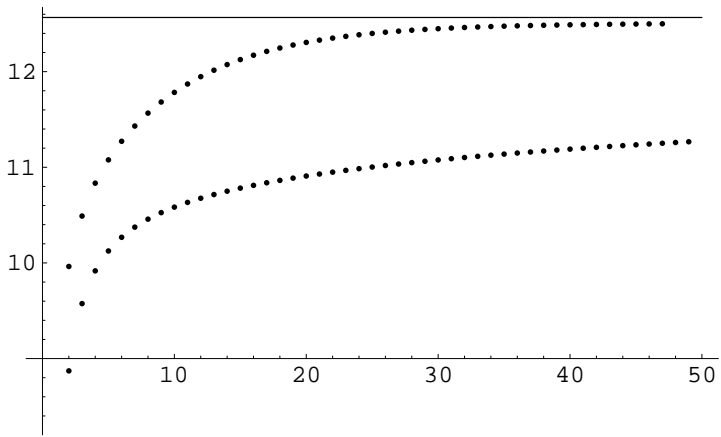, width=9.5cm}
\end{center}
\caption{Microscopic data for $f(N)=S(N)/ {\sqrt {N}}$ (bottom) and its transform $A(N)$ (top), defined in (\ref{rstu}), for the STU model, and for $1\le N<50$. The horizontal line 
is the expected asymptotic value $4\pi$.} \label{stu}
\end{figure}

The leading behavior (\ref{microentropy}) can be also verified by a numerical analysis similar to the one 
performed in sections 3 and 4. For example, for the STU model we have computed the quantity $f(N)=S(N)/ {\sqrt {N}}$ 
for $1\le N <50$, where $S(N)=\log\, \Omega(N)$. In order to subtract the 
logarithmic term in the asymptotic expansion (\ref{f6}) we consider the transform, 
\be
\label{rstu}
A(N)= { (N+1)S(N+2) -(2N+1) S(N+1)  + N S(N) \over 
(N+1){\sqrt {N+2}} -(2N+1) {\sqrt {N+1}} + N {\sqrt {N}} }. 
\ee
In \figref{stu} we plot $f(N)$ (bottom) and $A(N)$ (top). The horizontal 
line is the expected asymptotic value $4\pi$ for both quantities as $N \rightarrow \infty$. As before, the 
transform $A(N)$ improves rapidly the convergence.  

\subsection{Macroscopic entropy for small black holes}

The 5d black holes obtained by wrapping the M2 branes along cycles in the K3 fiber have actually vanishing
classical entropy and are therefore {\it small} black holes. Indeed, as we have seen, the leading asymptotic degeneracy scales
like $Q$, and not like $Q^{3/2}$. This is also what is found for small 4d black holes \cite{ddmp}.

Let us briefly show that the classical area of these black holes is zero for any set of intersection numbers $C_{abc}$, $C_{ab}$. In order to do this,
we can use the 5d attractor mechanism described in section 2. Equivalently, by using the 4d/5d connection of \cite{gsy}, we can
map the 5d black hole to a 4d black hole with D6 charge $p^0=1$ and D2 charges $Q_A$. At the level of the leading macroscopic entropy, the
4d computation gives the same result as the 5d computation \cite{gsy}. In the 4d language, we start with the tree level SUGRA prepotential
\be\label{prepo}
F=-\frac{1}{2}C_{ab}\frac{X^S X^a X^b}{X^0}-\frac{1}{6}C_{abc}\frac{X^a X^b X^c}{X^0}.
\ee
We will do the computation for a generic D6-D2 charge, i.e. we will start with generic charges $p^0$, $Q_a$, $Q_S$, and then
take the charge $Q_S\rightarrow 0$ at the end of the
computation (as well as setting $p^0=1$). This will guarantee that we obtain generic solutions to the
attractor mechanism.

Let us first assume that $C_{abc}=0$, as it happens in K3$\times \IT^2$ and the Enriques Calabi--Yau. In this case, the attractor equations are easily solved as,
\be\label{atstar}
(X^{0}_{*},X^{S}_{*},X^{a}_{*}) = \biggl( p^0,\ri \sqrt{\frac{p^0 Q^2}{2 Q_S}},\ri \sqrt{ {2 Q_S p^0 \over Q^2}}Q^a\biggr),
\ee
where
\be
Q^2 =C^{ab}Q_a Q_b, \qquad Q^a=C^{ab}Q_b.
\ee
The entropy is given by
\be
\label{smallentropy}
S=\pi \sqrt{2 p^0 Q_S Q^2},
\ee
and it vanishes in the limit $Q_S \rightarrow 0$. This is as expected.

If we now consider a general prepotential with nonvanishing $C_{abc}$, the attractor equations are now solved at
\be\label{Xy}
(X^0,X^S,X^a) = \biggl(p^0\;,\ri \sqrt{\frac{p^0}{2 Q_S}} \xi^S\;,\ri \sqrt{2 p^0 Q_S} \xi^a\biggr),
\ee
where the $\xi^A$ are solutions to,
\be
\label{equ1}
\ba
\xi^a \xi_a &= 1,\\
Q^a &= \xi^S \xi^a + Q_S C^{ab}C_{bef}\xi^e \xi^f.
\ea
\ee
Notice that, in these variables, the model with $C_{abc}=0$ corresponds to the smooth values,
\be
(\xi_{*}^S\;,\;\xi_{*}^a)\;=\Bigr({\sqrt{Q^2}}, {Q^a \over {\sqrt {Q^2}}}\Bigl).
\ee
We can already see that, in the limit $Q_S\rightarrow 0$, the perturbation by $C_{abc}$ in (\ref{equ1}) vanishes, therefore
in the limit of zero charge in the base the presence of nontrivial intersection numbers in the fiber should be unimportant. More formally, it is easy
to see that one can construct
a consistent solution of (\ref{equ1}) of the form,
\be
\xi^A=\xi^A_* + \sum_{n=1}^{\infty} c^A_n Q_S^n,
\ee
where the coefficients $c^A_n$ depend on $C_{abc}$ and can be calculated order by order. In terms of the $\xi^A$ the macroscopic entropy is 
\be
S=\pi \sqrt{2 p^0 Q_S} \biggl( C_{ab}  \xi^a \xi^b \xi^S+ \frac{2}{3}Q_S C_{abc}\xi^a \xi^b \xi^c \biggr),
\ee
and, in the limit
$Q_S \rightarrow 0$, it will vanish irrespectively of the value of $C_{abc}$. Therefore, 5d black holes whose membrane charge
is restricted to the K3 fiber of a K3 fibration are always small. This can be checked as well by detailed computations in different models
(like the STU and ST models considered above).

Since the leading contribution to the entropy vanishes we should now look at the subleading terms in the macroscopic entropy. As we explained in section 2, it
was shown in \cite{Guica:2005ig,Alishahiha:2007nn,Castro:2007hc} that these terms are obtained by performing the shift
\be
\label{Qshifted}
Q_A \rightarrow \widehat Q_A=Q_A + \zeta c_{2A}, \qquad \zeta={1\over 8}.
\ee
The leading term in the entropy for the small 5d black hole is given (for large charge $Q$) by performing this
shift in (\ref{smallentropy})
\be
\label{sublead}
S=2 \pi {\sqrt { {\zeta c_S \over 2} Q^2}}.
\ee
This can be derived in detail by solving the attractor equations with shifted charges (\ref{Qshifted}) as a power series in $1/Q$, and then taking the limit
$Q_S \rightarrow 0$. Notice that the entropy (\ref{sublead}) only depends on $C_{ab}$ and $c_S$. Also, in this regime, the solutions of the attractor equations occur at values of the
K\"ahler parameters which are of the order of the string size, and the SUGRA calculation might be problematic. Indeed, it is easy to see that (\ref{sublead})
does {\it not} agree with the leading term of the asymptotics that we obtained in the previous subsection. By comparing (\ref{microentropy}) with (\ref{sublead})
we find that the formula agree if we set instead $\zeta=1/6$. This is the value of $\zeta$ that is predicted by the 4d/5d connection of
\cite{gsy}.

In \cite{Guica:2005ig,Castro:2007hc} it was noticed that the subleading correction (\ref{Qshifted}) obtained in a macroscopic
5d computation was not in accord with the subleading correction predicted by \cite{gsy} and the 4d attractor mechanism.
We now find that, for {\it big} 5d black holes, the subleading correction for the microscopic entropy is in rough agreement with (\ref{Qshifted}), while
for {\it small} 5d black holes the leading asymptotics is in accord with a 4d computation for a small D6/D2 system with $p^0=1$. As we already mentioned,
in the case of small black holes, the SUGRA computations with which we are comparing our results
should receive large corrections, but in other situations they still lead to results which are in 
agreement with the microscopic counting, as in \cite{ddmp,df}. In our case we obtain a result in 
disagreement with the 5d computation but in agreement with the 4d computation. It would be 
interesting to resolve this puzzle.

\section{Conclusions}
In this paper we have studied the microscopic counting of 5d black hole states
by using topological string theory. In the case of big black holes, we have given convincing numerical evidence
that the BPS invariants encoded in the topological string amplitudes account correctly for the
macroscopic entropy of spinning black holes. Moreover, we have also shown that the data 
favour the ``mysterious cancellation"
of \cite{dm} that makes possible to extend the validity of the OSV conjecture, and we were able to
explore new aspects of black hole entropy which have not been studied
before using supergravity. Clearly, it would be very desirable to improve our numerical results with more
data. Using the interplay between modularity and an-holomophicity in topological 
string theory~\cite{Yamaguchi:2004bt,hkq,gkmw}, analytic results on the asymtotics might be not out 
of reach\footnote{Recently beautiful analytic proofs of the asymptotic of the Fourier coefficents 
of Mock-Theta functions have been obtained using a somewhat similar interplay~\cite{KO}.}.

We also gave exact formulae for microscopic degeneracies of
a class of small 5d black holes, which are obtained by wrapping M2 branes in the fiber of a K3 fibration, 
and we computed the asymptotic expansion in inverse powers of the charge. As expected, 
the calculation shows that for small black holes the leading term in the entropy scales like $S\rightarrow \lambda S$ when the 
charges are scaled with $\lambda$. We found however that the coefficient of the leading term does not agree with 
the shift of charges obtained in \cite{Guica:2005ig,Alishahiha:2007nn,Castro:2007hc} in a 5d SUGRA computation. 
In principle there is no reason why these two computations should agree, since small $\CN=1$ black holes are 
generically beyond the SUGRA approximation. On the other hand, the microscopic results are well reproduced by 
the 4d/5d connection of \cite{gsy} and a 4d attractor computation. We should emphasize however that for big 
black holes the 5d shift (\ref{Qshifted}) fits our data better than the 4d shift with $\zeta=1/6$. 
It would be very interesting to understand this better.

\section*{Acknowledgments}
It is a pleasure to thank Davide Gaiotto, Thomas Grimm, Aki Hashimoto, Sheldon Katz, Wei Li, Boris Pioline, 
Nick Warner and Xi Yin for helpful
discussions, and Frederik Denef for a very useful correspondence. We would like to thank as well Gregory Moore and Cumrun Vafa 
for their comments on the manuscript. Many thanks also to Max Kreuzer for generously granting us computer
time. This work is partially supported by the DOE grant DE-F602-95ER40896. AT is supported by a Marie Curie fellowship.

\appendix

\section{General features of the instanton expansion} \label{instantons}

The asymptotic behaviour at the conifold, Castelnouvo's theory, and the 
calculation via degenerate Jacobians, suggest some general features of 
the Gopakumar--Vafa expansion. Our data for the $13$ one-parameter models 
suggest further universal features. The purpose of this appendix is 
to describe some of these general features. Typical data for high degree 
look as is table \ref{33degree18} 

\begin{table}
\begin{centering}
\begin{tabular}{|r|c|}
\hline
genus & degree=18 \\
\hline
0&  144519433563613558831955702896560953425168536  \\
1&  491072999366775380563679351560645501635639768 \\
2&  826174252151264912119312534610591771196950790 \\
3&  866926806132431852753964702674971915498281822\\
4&  615435297199681525899637421881792737142210818\\
5&   306990865721034647278623907242165669760227036 \\
6&   109595627988957833331561270319881002336580306 \\
7&   28194037369451582477359532618813777554049181 \\
8&   5218039400008253051676616144507889426439522 \\
9&   688420182008315508949294448691625391986722 \\
10&  63643238054805218781380099115461663133366  \\
11&  4014173958414661941560901089814730394394  \\
12&   166042973567223836846220100958626775040 \\
13&  4251016225583560366557404369102516880  \\
14&   61866623134961248577174813332459314 \\
15&   451921104578426954609500841974284 \\
16&   1376282769657332936819380514604 \\
17&   1186440856873180536456549027 \\
18&   2671678502308714457564208 \\
19&   -59940727111744696730418 \\
20&   1071660810859451933436 \\
21&   -13279442359884883893 \\
22&   101088966935254518, \\
23&    -372702765685392\\
24&    338860808028\\
25&   23305068 \\
26&   -120186 \\
27&   -5220 \\
28&   -90 \\
29&   0 \\
 \hline
\end{tabular}  \caption{\label{33degree18} Gopakumar--Vafa invariants $n_d^g$ 
in the class $d=18$ for the complete intersection $X_{3,3}(1^6)$.}
\end{centering}
\end{table}

The last nonzero entry is from the smooth genus $28$ complete intersection 
curve\footnote{A complete intersection curve $(1,n,3,3)$ with degree $9n$ has in general 
genus $\tilde g=\frac{1}{2}( 1+ 3 n) (2 + 3 n)$.}    $(1,2,3,3)$ of degree $18$. 
By Castelnouvo's theory $\tilde g=28$ is the largest possible genus for 
degree $18$.   The degree one constraint parametrizes an 
$\mathbb{P}^5$. The moduli space ${\cal M}_{18}^{28}$ is a fibration of 
this  $\mathbb{P}^5$ over a projectivization of the $15$ parameters in 
the quadratic constraint. I.e. ${\cal M}_{18}^{28}$ is the total space of 
$\mathbb{P}^5\rightarrow \mathbb{P}^{14}$, 
with Euler number $\chi( {\cal M}_{d=18}^{g=28})=5\times 15 =90$ and $n_{18}^{28}
=(-1)^{5+14} 90=-90$. 

\begin{figure}
\begin{center}
\epsfig{file=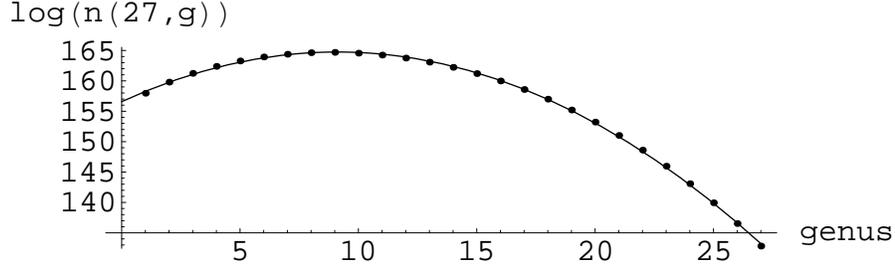,width=12cm}
\end{center}
\caption{The binomials dominate the behaviour of large $d$ Gopakumar-Vafa 
invariants. For the degree 27 class  on the bi-cubic we find 
$n^g_{27}\sim e^{167.747} e^{-0.0985 (g-9.108)^2}$ }
\label{binomial}
\end{figure}

As it can further be seen in table \ref{33degree18},  the numbers grow from genus 
$g=0$ to $g=3$ and fall thereafter. This feature might be related to the binomials in 
the description of the moduli of space as a singular fibration of the Jacobian 
${\rm Jac}_{28}$ of the $g=28$ curve over ${\cal M}_{18}^{28}$. In 
this description  the contribution of a $g=\tilde g-\delta$ curve 
comes from degenerating the genus $28$ curve with $\delta$ nodes. As 
explained in \cite{kkv} the contribution of the degenerate Jacobians 
can be expressed by the Euler numbers of relative Hilbert schemes 
${\cal C}^{(n)}$ as 
\begin{equation}
n^{\tilde g-\delta}_d=(-1)^{{\rm dim}({\cal M})+\delta} \sum_{p=0}^\delta b(\tilde g-p,\delta-p)
\chi({\cal C}^{(n)})\ , 
\label{hilbertscheme}
\end{equation} 
with $b(g,k)=\left(2 g-2 \atop k\right)$.  
A simple Gauss approximation of binomials fits the behaviour of the $n^g_d$ for large 
$d$ relatively well. We show this in Fig. \ref{binomial} for the bi-cubic at degree $27$. 
The numbers $n^g_d$ are exact and in contrast to (\ref{hilbertscheme}) they  count 
correctly all contribution from colliding nodes, all contributions from 
reducible curves as well as contributionsfrom smooth curves 
in the class $d$ with genus $\tilde {\tilde g} < \tilde g$.

\begin{figure}[hbtp]
\begin{center}
\epsfig{file=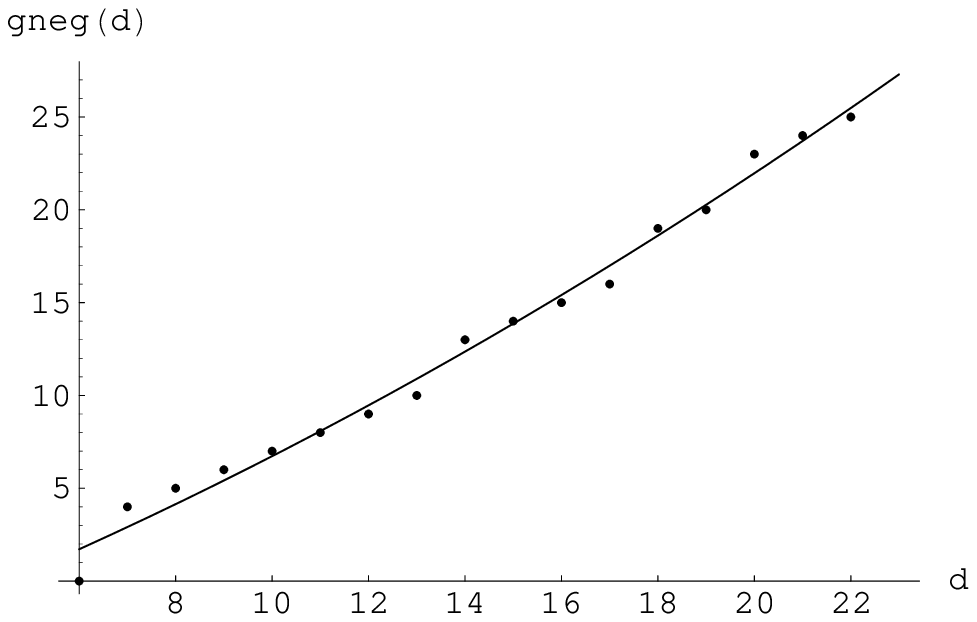, width=8cm}\epsfig{file=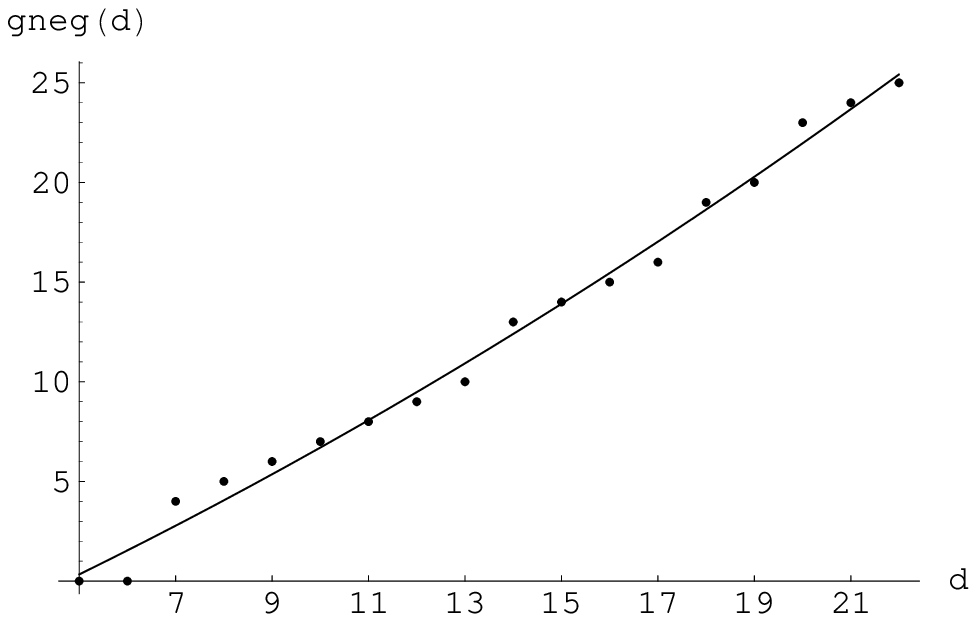, width=8cm}
\end{center}
\caption{The first occurrence of negative $n^g_d$ for the quintic (on the right) and the bi-cubic (on the 
left). The fit is $m(d)=a+ bd +c d^2$ with $a=-4.6$, $b=.94$ and $c=.019$ as well as  
$a=-5.2$, $b=1.0$ and $c=.017$ for these two , respectively.} 
\label{negative}
\end{figure}

Very important for the cancellations in the asymptotic behaviour of the Donaldson--Thomas 
invariants  is the occurrence of negative numbers. While it is clear that such contributions
can arise if the dimensions of the D-brane moduli space is odd, we do not understand a priori 
the remarkable pattern with which these signs occur. The first occurrence of negative signs at 
$g_{neg}(d)$  is graphed for the quintic and the bi-cubic in \figref{negative}. The data 
suggest that $g_{neg}(d)$ follows a parabola similar to the Castelnouvo bound. From the first 
occurrence of the negative sign the $n^g_d$ are alternating in sign for $g\ll \tilde g$. 
For  $g\sim \tilde g$ the behaviour becomes more erratic. The Gauss approximation for the absolute 
values of the $n^g_d$ and the sign pattern is very characteristic of the degeneracies of 
microstates of a large black hole. In contrast the absolute value of the $n^g_d$ is falling and the   
signs are alternating with $(-1)^g$ starting at $g=0$ for small black holes as shown for 
example for the ST-model.          

\vskip 5 mm {\vbox{\small{
$$
\vbox{\offinterlineskip\tabskip=0pt \halign{\strut \vrule#& 
&\hfil~$#$ 
&\hfil~$#$ 
&\hfil~$#$ 
&\hfil~$#$ 
&\hfil~$#$
&\hfil ~$#$
&\hfil ~$#$
&\hfil ~$#$
&\vrule#\cr \noalign{\hrule} 
&g&d=1&2         &3          &4       &5       &6       &7 &\cr
\noalign{\hrule}
&0&2496&223752&38637504&9100224984&2557481027520&805628041231176&\ldots&\cr 
&1&0&-492&-1465984&-1042943520&-595277880960&-316194812546140&\ldots&\cr 
&2&0&-6&7488&50181180&72485905344&70378651228338&\ldots&\cr 
&3&0&0&0&-902328&-5359699200&-10869145571844&\ldots&\cr 
&4&0&0&0&1164&228623232&1208179411278&\ldots& \cr 
&5&0&0&0&12&-4527744&-94913775180&\ldots& \cr 
&6&0&0&0&0&17472&4964693862&\ldots& \cr 
&7&0&0&0&0&0&-152682820&\ldots&\cr
&8&0&0&0&0&0&2051118&\ldots&\cr 
&9&0&0&0&0&0&-2124&\ldots&\cr
&10&0&0&0&0&0&-22&605915136&\cr
&11&0&0&0&0&0&0&-9419904& \cr
&12&0&0&0&0&0&0&32448&\cr
\noalign{\hrule}}\hrule}$$}}} 
                   
\begin{figure}[hbtp]
\begin{center}
\epsfig{file=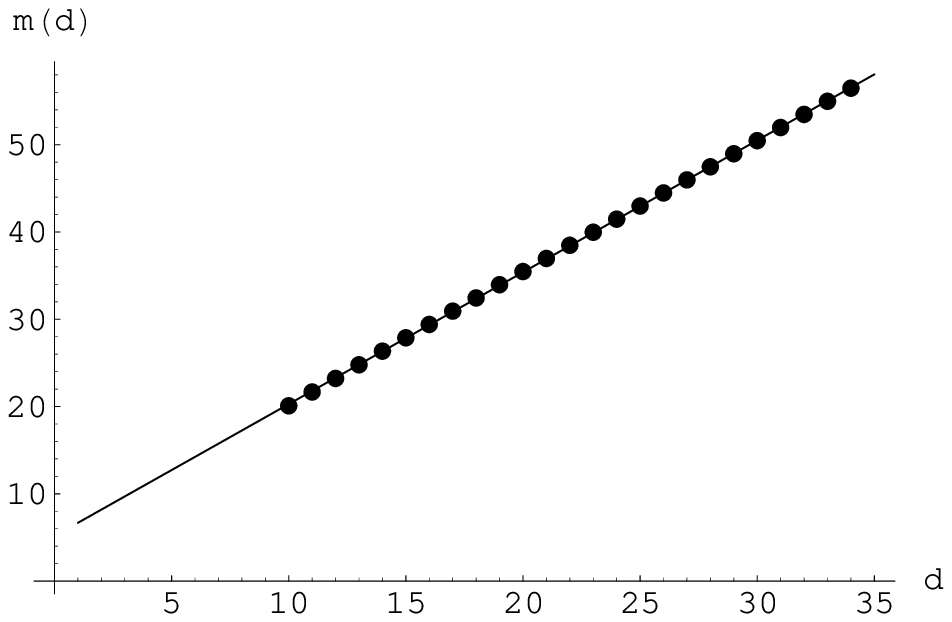, width=8cm}\epsfig{file=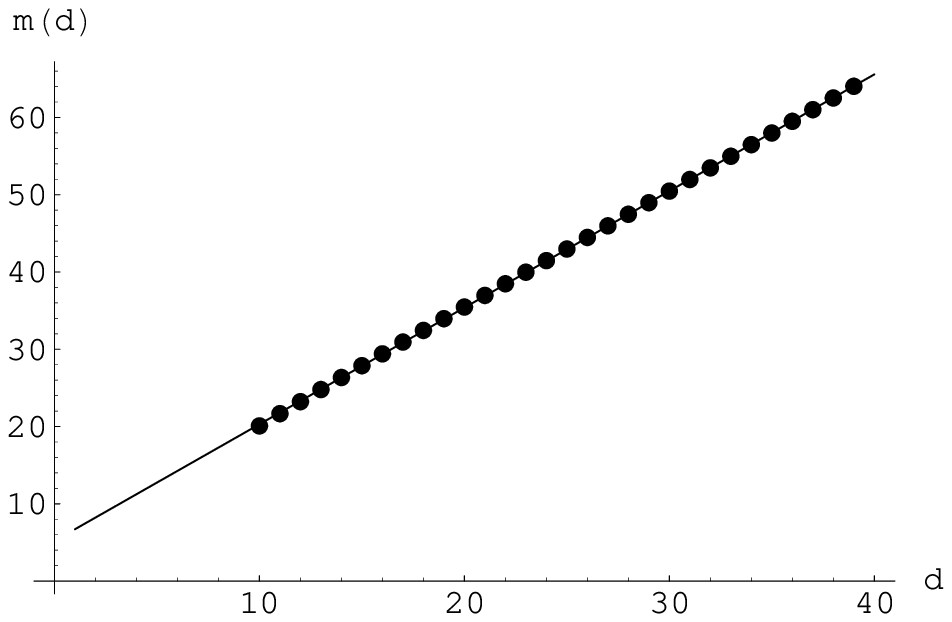, width=8cm}
\end{center}
\caption{$m(d)=\log(M(d))^{3/4}$ for the quintic on the right and the bi-cubic on the 
left.  $a=5.164$ and  $b=1.511$ as well as  $a=5.202$ and  $b=1.509$ for the cases plotted.} 
\label{scaling}
\end{figure}

A further remarkable fact is the very universal scaling for the 
maximal value $M(d)$ for  
$n^g_d$ for given $d$. This value behaves like
\begin{equation} 
M(d)=\exp\left((a+bd)^{4/3}\right)\ 
\end{equation}
with very  similar values for $a$ and $b$ for different one-parameter models, as shown for the quintic and 
the bi-cubic in \figref{scaling}.


\end{document}